\documentclass[12pt]{article}
\usepackage{geometry}
\usepackage{amsmath, amssymb, amsthm}
\usepackage{float}

\usepackage{graphicx}
\usepackage[font=footnotesize]{caption}
\usepackage{subfigure} 
\usepackage[square,numbers]{natbib}
\usepackage{booktabs}
\usepackage[utf8]{inputenc}
\usepackage{amsmath}   
\usepackage{amssymb}   
\usepackage{array}  
\usepackage{hyperref}

\usepackage{booktabs}  
\usepackage{lipsum}    
\usepackage{array}       
\usepackage{multirow}    
\usepackage{makecell}    
\usepackage{amsmath}     
\usepackage{nicematrix}

\usepackage{subfiles}

\usepackage{xr}
\usepackage{cleveref}
\makeatletter
\newcommand*{\addFileDependency}[1]{
\typeout{(#1)}
\@addtofilelist{#1}
\IfFileExists{#1}{}{\typeout{No file #1.}}
}\makeatother

\newcommand*{\myexternaldocument}[1]{%
\externaldocument{#1}%
\addFileDependency{#1.tex}%
\addFileDependency{#1.aux}%
}
\myexternaldocument{supplementary_Information}

\usepackage{media9}
\usepackage{multimedia}

  \title{Pollutant-induced changes in fish pigmentation and spatial patterns}
 
\author{
Pranali Roy Chowdhury$^{1,3,\dagger}$,
Tian Xu Wang$^{1,\dagger}$,
Abbey MacDonald$^{2}$,\\
Keith B.\ Tierney$^{2}$,
Hao Wang$^{1,\ast}$
}

\date{}

\begin{document}

\maketitle
\noindent{} 1. Department of Mathematical \& Statistical Sciences, University of Alberta, Edmonton, Canada\\
\noindent{} 2. Department of Biological Sciences, University of Alberta, Edmonton, Canada\\
\noindent{} 3. Department of Computational \& Data Sciences, Indian Institute of Science Bengaluru, India

\bigskip

\noindent{} $\dagger$ These authors contributed equally to this work.\\
\noindent{} $\ast$ Corresponding author; e-mail: hao8@ualberta.ca.

\bigskip

\begin{center}
{\bf Abstract}
\end{center}

Pigmentation abnormalities, ranging from hypo- to hyperpigmentation, can serve as biomarkers of developmental disruption in fish exposed to environmental contaminants. However, the mechanistic pathways underlying these alterations remain poorly understood. Studies have shown that pattern formation in fish development requires specific pigment cell interactions. Motivated by experimental observations of pigmentation alterations following contaminant exposure, we investigate how pollutants influence pigment cell self-organization using a continuum reaction–diffusion–advection framework. The model incorporates nonlocal Morse-type kernels to describe short- and long-range interactions among melanophores and xanthophores. Our results show that perturbations to the strengths of adhesion or repulsion can drive transitions between stripes, spots, and mixed patterns, reproducing phenotypes characteristic of fish pigmentation mutants. In particular, homotypic interactions are sensitive to contamination, leading to pronounced changes in melanophore density and resulting pigmentation patterns. Time-dependent simulations indicate that pigment changes from early short-term contaminant exposure may be recoverable, whereas prolonged exposure can lead to sustained pigment loss. In a growing fish, contaminant-induced changes in cell–cell interactions directly influence stripe formation rate, stripe number, and pigmentation levels. Overall, our study provides insight into the mechanistic link between experimentally observed pigmentation alterations and the changes in spatial patterns of adult fish.

\vspace{1.0cm}
\noindent
{\bf Keywords:} Hypopigmentation; Pattern formation; Nonlocal diffusion-advection; Water pollutants; Pigment cell interaction.

\maketitle

\newpage
\section{Introduction}

\begin{figure}[ht!]
    \centering
    \includegraphics[width=8cm,height=6cm]{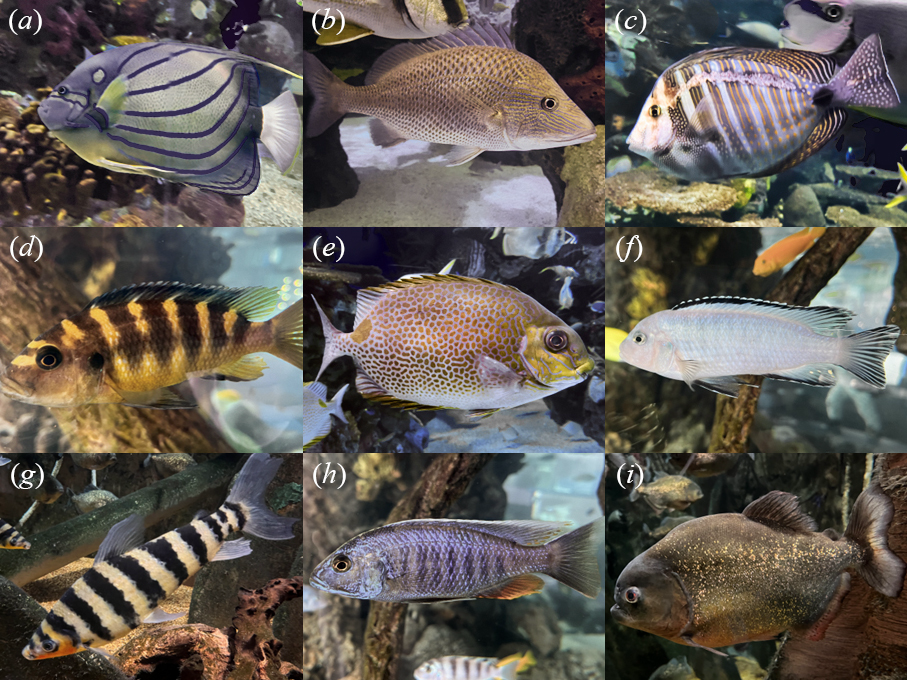}
    \caption{Examples of marine and freshwater species displaying a wide variety of pigment patterning outcomes. These patterns are determined by the number, spatial arrangement, and interactions of melanophores, xanthophores, iridophores, and other chromatophore classes. (a) Blue-ring angelfish \textit{Pomacanthus annularis}; (b) emperor bream (genus \textit{Lethrinus}); (c) Desjardini sailfin tang \textit{Zebrasoma desjardinii}; (d) bumblebee cichlid \textit{Pseudotropheus crabro}; (e) rabbitfish (genus \textit{Siganus}); (f) pale African cichlid; (g) banded leporinus \textit{Leporinus fasciatus}; (h) electric blue hap \textit{Sciaenochromis fryeri}; (i) red-bellied piranha \textit{Pygocentrus nattereri}.}
    \label{fig:real fish pattern}
\end{figure}

Fish colour patterns not only provide protection from predators but also serve as important recognition signals in social communication \cite{walderich2016homotypic}. These patterns arise from the spatial organization and interactions of pigment cells, and their formation has been extensively studied in fish \cite{dawid2004developmental,Volkening2015,Nakamasu2009}. For example, the characteristic black-and-yellow striped patterns emerge from coordinated interactions among three major pigment cell types: melanophores, xanthophores, and iridophores \cite{Nakamasu2009,Martinson2024,Kondo2021,haffter1996mutations}. These pigment cells can give rise to an incredible diversity of pigment patterns (Figure~\ref{fig:real fish pattern}).

Pigmentation abnormalities in fish can be triggered by a wide range of environmental pollutants. As anthropogenic activities intensify worldwide, water pollution has become a major global concern \cite{jones2023sub}, with aquatic ecosystems increasingly acting as sinks for diverse contaminants \cite{bashir2020concerns}. Major sources of water pollution include agricultural runoff, industrial discharges, and sewage effluents \cite{haddeland2014global,ejiohuo2024ensuring}, which introduce a variety of pollutants such as heavy metals (e.g., mercury, lead, cadmium), excess nutrients (e.g., nitrates, phosphates), industrial organic compounds (e.g., pesticides and PAHs), pharmaceuticals, and microplastics. Exposure to these pollutants can alter fish pigmentation, resulting in either hypopigmentation or hyperpigmentation. Hypopigmentation is typically associated with delayed development, impaired growth, or pigment cell dysfunction, leading to reduced melanin production and lighter body coloration, whereas hyperpigmentation is associated with accelerated development or enhanced pigment cell activity, resulting in darker pigmentation. Several compounds have been reported to stimulate pigmentation. For example, fisetin, a natural flavonoid found in fruits and vegetables such as grapes and onions, increases both intracellular and extracellular melanin levels in B16F10 cells and enhances melanogenesis in zebrafish (\emph{Danio rerio}) larvae \cite{molagoda2020gsk}. Similarly, flumequine, a second-generation quinolone antibiotic known to induce phototoxicity, and rice bran ash mineral extract have been shown to stimulate pigmentation in zebrafish by upregulating key transcriptional pathways involved in melanin biosynthesis and transport \cite{karunarathne2019flumequine,kim2019rice}. In contrast, several contaminants have also been identified that induce hypopigmentation in zebrafish, including carbendazim and cadmium \cite{Schmidt2016,zhang2015mechanisms}, halogenated dipeptides \cite{peng2020evaluation}, and complex wastewater mixtures containing plasticisers, propellant precursors, and aromatic compounds \cite{mu2020pigmentation}.

Proper fish pattern development requires the coordinated specification and differentiation of pigment cells, along with precise cellular interactions \cite{patterson2019zebrafish}. The absence of any one pigment-cell type disrupts the formation of normal stripe patterns. Over the past few decades, numerous experimental and theoretical studies have sought to uncover the mechanisms of pigment-cell interactions responsible for body pattern formation in fish \cite{Nakamasu2009,Kondo2021,asai1999zebrafish,kondo2010reaction,painter1999stripe,yamaguchi2007pattern,watanabe2015pigment,Volkening2015}. These cellular interactions are believed to involve the ability of pigment cells to recognize one another and aggregate \cite{maderspacher2003formation}. Using in vitro analyses, Yamanaka et al. \cite{yamanaka2014vitro} showed that adult pigment patterns are driven by direct, contact-dependent interactions between pigment cells. Such interactions can occur over short or long ranges, either through direct cell–cell contact or via protrusive structures such as pseudopodia. Ablation experiments have further confirmed that pigmentation patterns can regenerate, supporting the concept of an autonomous mechanism governed by “short-range activation and long-range inhibition” \cite{Nakamasu2009}.

Pollutant-induced changes in fish pigmentation are thought to occur primarily through disruptions of pigment cell–cell interactions. Many environmental contaminants can interfere with normal cellular processes by interfering with these interactions. For example, microplastics can adhere to cell surfaces through adsorption and aggregation \cite{zhang2017toxic} and may compromise cell walls via surface absorption \cite{liu2019microplastic}, thereby affecting pigment formation and cell–cell communication. Similarly, widely distributed chemicals such as Bisphenol A (BPA) and Bisphenol F (BPF) can alter melanin synthesis in zebrafish embryos \cite{mu2020pigmentation}, with BPA disrupting integrin signaling and lipid raft domains that are critical for proper cell adhesion and signaling \cite{izard2025environmental}. Heavy metals, such as lead (Pb), can further compromise cellular integrity by altering the molecular composition of proteins, lipids, and DNA, leading to protein degeneration and lipid peroxidation \cite{huang2012tissue}. Moreover, some endocrine-disrupting chemicals can modify gene expression in immune cells, affecting cell–cell interactions and overall cellular communication networks \cite{das2025modulation}. Overall, data suggests a diversity of pollutants may alter pigment via cell interactions, ultimately disrupting the normal pattern formation in fish.

Theoretical modelling provides a powerful framework for qualitatively and quantitatively exploring pigment cell interactions. Since Alan Turing proposed a hypothetical mechanism in 1952 to explain patterns observed in animal skin \cite{Turing1952}, numerous theoretical studies have focused on cell interactions using continuum partial differential equation (PDE) models, lattice-based cellular automata, and agent-based modelling approaches \cite{Volkening2015,painter2015nonlocal,painter1999stripe}. However, to date, no studies have specifically examined the effects of environmental contaminants on pigment cell interactions. Understanding how water contaminants influence developmental processes in aquatic organisms requires a mechanistic framework that links observed biological patterns with underlying cellular and physical mechanisms. To address this gap, we incorporate pollutants into a model of pigment cell interactions to examine how contaminants affect cell–cell communication, proliferation, migration, and short- or long-range signaling. In particular, we formulate a general reaction–diffusion–advection model to investigate hypo- and hyperpigmentation under varying levels of water pollutants.

Most chemical pollutants in aquatic environments rarely reach lethal levels, highlighting the importance of studying sub-lethal effects to capture the full spectrum of adverse chemical impacts on aquatic species. Sub-lethal exposures are expected to induce subtle changes, such as in cell interactions, rather than cause cell death. From a mathematical perspective, our approach allows us to identify which pigment cell interactions are most sensitive to pollutant exposure and most likely to be disrupted, potentially giving rise to mutant pigmentation patterns \cite{mcguirl2020topological}. We also conducted complementary experiments involving methane exposure to further support these findings.



\section{Fish pattern formation driven by melanophore–\\xanthophore interactions}

Pigment cells can interact over both short and long ranges \cite{yamanaka2014vitro}. Short-range interactions involve direct cell–cell contact or communication mediated by short dendritic extensions. In contrast, long-range interactions are facilitated by melanophores extending elongated protrusions, known as pseudopodia, which transmit signals to cells located at a longer distance \cite{kondo2012turing}. Contact-dependent cell adhesion promotes cell–cell attachment and aggregation, whereas cell–cell repulsion prevents overcrowding and maintains appropriate intercellular spacing, thereby sharpening the boundaries of pigment patterns. Although several pigment cell types contribute to fish coloration, interactions between melanophores and xanthophores alone are sufficient to generate self-organized body patterns \cite{Kondo2021}; these two cell types are the focus of this study. Genetic analyses have revealed the presence of homotypic interactions among melanophores and xanthophores \cite{maderspacher2003formation}. Homotypic interactions, occurring between cells of the same type (melanophore–melanophore and xanthophore–xanthophore) play a crucial role in pigment cell proliferation, dispersal, and tiling within the skin \cite{walderich2016homotypic}. By regulating intercellular spacing, these interactions ensure a well-distributed layer of each pigment cell type across the fish body.

\begin{figure}[ht!]
    \centering
    \includegraphics[width=11cm,height=8cm]{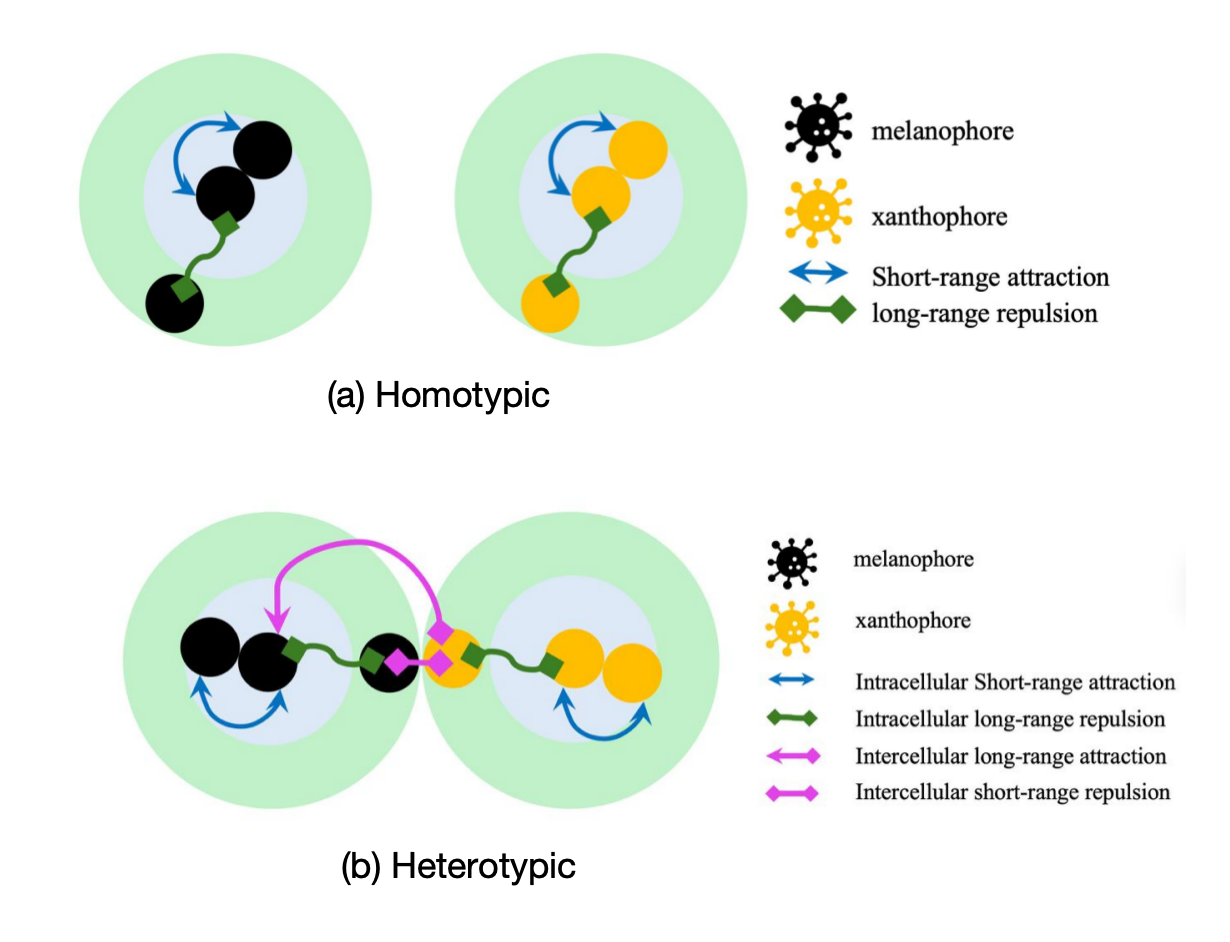}%
    \caption{Illustrations of homotypic and heterotypic interactions. 
(a) In the homotypic case, only short-range attraction and long-range repulsion occur among cells of the same type, with no interactions between different cell types. 
(b) In the heterotypic case, an additional one-directional long-range attraction and a short-range repulsive interaction between the two cell types are present, producing a chase–and–run–type behavior in which xanthophores chase melanophores while melanophores avoid xanthophores.}
\label{fig:Illustrations of homotypic and heterotypic interactions}
\end{figure}

 Heterotypic interactions occur between melanophores and xanthophores. Xanthophores extend pseudopodia toward melanophores, inducing melanophore movement away from the contact site, while xanthophores actively pursue them (Fig.~\ref{fig:Illustrations of homotypic and heterotypic interactions}), giving rise to a “run-and-chase” behavior. Heterotypic interactions strongly influence pigment cell morphology, and homotypic interactions primarily regulate cell proliferation and dispersal \cite{walderich2016homotypic}. Together, these coordinated interactions generate the distinct spatial organization of pigment cells that underlies fish color pattern formation.



\subsection{Modelling pigment cell interactions for pattern formation}
\label{sec:model formulation}

 Let \(u(x,t)\) and \(v(x,t)\) denote the densities of melanophores and xanthophores, respectively. We approximate the fish surface as a two-dimensional spatial domain \(\Omega \subset \mathbb{R}^2\). These cells interact through migration, proliferation, and death. Migration is modeled as a combination of random diffusion and directed, nonlocal advection driven by cell--cell interactions, while proliferation and death are described by local interaction terms \(f(u,v)\) and \(g(u,v)\). The resulting reaction--diffusion--advection system is
\begin{equation}\label{mod:Movement_model}
    \begin{aligned}
        u_t & = D_u \Delta u 
        - D_{uv}\nabla \cdot \bigl(u \nabla (G_{uv} * v)\bigr) 
        - D_{uu}\nabla \cdot \bigl(u \nabla (G_{uu} * u)\bigr) 
        + f(u,v), \quad \text{in } \Omega \times (0,T),\\
        v_t & = D_v \Delta v 
        - D_{vu}\nabla \cdot \bigl(v \nabla (G_{vu} * u)\bigr) 
        - D_{vv}\nabla \cdot \bigl(v \nabla (G_{vv} * v)\bigr) 
        + g(u,v), \quad \text{in } \Omega \times (0,T).
    \end{aligned}
\end{equation}
Here, \(D_u\) and \(D_v\) denote the diffusion coefficients of melanophores and xanthophores, respectively. The directed movement is governed by interaction kernels \(G_{ij}\), chosen as truncated Morse potentials to capture both attractive and repulsive forces within a finite interaction range \cite{bhaskar2023topological}:
\[
   G_{ij}(x,y) =
   \begin{cases}
   A_{ij} e^{-(x^2+y^2)/d_{ij}^a}
   - R_{ij} e^{-(x^2+y^2)/d_{ij}^r}, & (x^2+y^2) \le d_a,\\
   0, & \text{otherwise},
   \end{cases}
\]
where \(i,j \in \{u,v\}\). Each kernel is smooth and has compact support within a disk of radius \(d_a\), reflecting the maximum distance over which cells interact directly or via protrusions such as pseudopodia. The parameters \(A_{ij}\) and \(R_{ij}\) denote the strengths of attraction and repulsion, while \(d_{ij}^a\) and \(d_{ij}^r\) represent the corresponding interaction length scales.

The convolution operator
\[
(G_{ij} * j)(x,y)
= \int_{\Omega} G_{ij}(x-x',\, y-y')\, j(x',y')\, dx' \, dy'
\]
describes the cumulative nonlocal influence of cells of type \(j\) on cells of type \(i\) at position \((x,y)\). The homotypic kernels \(G_{uu}\) and \(G_{vv}\) describe interactions between cells of the same type, whereas the heterotypic kernels \(G_{uv}\) and \(G_{vu}\) describe interactions between different cell types: \(G_{uv}\) captures the influence of xanthophores on melanophores, and \(G_{vu}\) captures the effect of melanophores on xanthophores. The coefficients \(D_{uu}, D_{uv}, D_{vu},\) and \(D_{vv}\) quantify the strength of these interaction-induced advective effects.

The advective terms \(\nabla \cdot (u \nabla (G_{uv} * v))\) and \(\nabla \cdot (u \nabla (G_{uu} * u))\) represent directed movement of melanophores in response to heterotypic and homotypic interactions, respectively. Depending on the balance between attractive and repulsive components of the interaction potentials, these forces can drive cells either toward or away from one another. The interplay of diffusion, nonlocal interactions, and local growth dynamics ultimately gives rise to self-organized pigment patterns on the fish body.

The well-posedness of the model has been extensively studied in the literature, including cases with less regular interaction kernels \cite{giunta2025phylogeny,Giunta2025Positivity}, and it has been shown that the system can generate a wide variety of patterns. For numerical simulations, we consider a static two-dimensional domain representing a portion of the fish surface, with periodic boundary conditions. Due to the complexity of system \eqref{mod:Movement_model}, it is generally not possible to determine the full parameter space from experimental data. Therefore, rather than exploring all possible emergent patterns, we focus specifically on spot and stripe formations.  

Diffusion parameters are chosen based on ranges previously used to simulate spot patterns in fish leopard mutants \cite{nakamasu2022correspondences}. The interaction strengths are set uniformly as \(D_{uu} = D_{uv} = D_{vu} = D_{vv} = 0.01\), while the random motility coefficients are \(D_u = D_v = 0.1\). To generate spot patterns, the system is initialized with small random perturbations around mean cell densities:
\begin{equation}\label{Eq:initial_cond}
    u_0 = \tilde{u} + \varepsilon_1 \xi_1(x,y), \quad v_0 = \tilde{v} + \varepsilon_2 \xi_2(x,y),
\end{equation}
where \(\xi_1\) and \(\xi_2\) are random fields. In line with \textit{in vivo} observations showing that xanthophores spread faster across the skin than melanophores \cite{walderich2016homotypic}, we assign a higher amplitude to xanthophore fluctuations (\(\varepsilon_2 > \varepsilon_1\)) to capture the spatial heterogeneity of cell distribution.

The pattern formation arises due to the attractive and repulsive force between the interacting cells. Exposure to toxic pollutants can disrupt these signaling mechanisms by altering cell kinetics, adhesion strength, and cell communications. In our system, we assume that the fish is being constantly exposed to the external pollutant, and this exposure can influence the interaction strength between melanophores and xanthophores. 

\subsection{Homotypic and heterotypic cell interactions}

We first consider homotypic interactions (Figure~\ref{fig:Illustrations of homotypic and heterotypic interactions}(a)) by setting $G_{uv} = G_{vu} = 0$, incorporating short-range attraction and long-range repulsion among cells of the same type, such that $d^a_{uu} < d^r_{uu}$ and $d^a_{vv} < d^r_{vv}$. In this case, the model \eqref{mod:Movement_model} accounts only for interactions between $u$–$u$ and $v$–$v$ cells:
\begin{equation}
\begin{aligned}
   G_{uu}  &= 
   \begin{cases}
      A_{uu} e^{-(x^2+y^2)/d_{uu}^a} - R_{uu} e^{-(x^2+y^2)/d_{uu}^r}, & (x^2+y^2) \le d_u^e,\\
      0, & \text{otherwise},
   \end{cases}\\
   G_{vv}  &= 
   \begin{cases}
      A_{vv} e^{-(x^2+y^2)/d_{vv}^a} - R_{vv} e^{-(x^2+y^2)/d_{vv}^r}, & (x^2+y^2) \le d_v^e,\\
      0, & \text{otherwise}.
   \end{cases} 
\end{aligned}
\end{equation}
Within this framework, we investigate how external pollutants perturb both homotypic and heterotypic interactions over short and long timescales, thereby influencing pigment pattern formation. The long-range repulsion parameters are set as $d_{uu}^r = 3\,d_{uu}^a$ and $d_{vv}^r = 3\,d_{vv}^a$, while effective cut-off distances are $d_u^{e} = 5\,d_{uu}^a$ and $d_v^{e} = 5\,d_{vv}^a$, beyond which cells no longer interact. Competitive interactions are incorporated via logistic-like reaction terms: 
\[
f(u,v) = u(1-u-v), \quad g(u,v) = v(1-u-v).
\]
Under these settings, the model produces Turing-like spot patterns, comparable to those observed in leopard mutant fish \cite{Singh2015}, with variations in spot size, density, and connectivity corresponding to different alleles \cite{asai1999zebrafish}. Thus, our reaction–diffusion–advection model \eqref{mod:Movement_model} effectively captures the formation of pigment patterns in fish.

We next consider heterotypic interactions between melanophores and xanthophores (Figure~\ref{fig:Illustrations of homotypic and heterotypic interactions}(b)). \textit{In vitro} experiments describe these interactions as directed movement dynamics \cite{Kondo2021,jewell2025chase,painter2024variations}. In the model, cell proliferation and death follow logistic growth: 
\[
f(u,v) = u(1-u), \quad g(u,v) = v(1-v).
\] 
Melanophores move away from xanthophores, while xanthophores actively pursue melanophores, giving rise to the characteristic “run-and-chase” behavior. We capture this asymmetry using short-range repulsion in both directions and one-directional long-range attraction:
\begin{equation}
\begin{aligned}
   G_{uv} &= 
   \begin{cases}
      A_{uv} e^{-(x^2+y^2)/d_{uv}^a} - R_{uv} e^{-(x^2+y^2)/d_{uv}^r}, & (x^2+y^2) \le d_{uv}^e,\\
      0, & \text{otherwise}, 
   \end{cases}\\[1mm]
   G_{vu} &= 
   \begin{cases}
      - R_{vu} e^{-(x^2+y^2)/d_{vu}^r}, & (x^2+y^2) \le d_{vu}^e,\\
      0, & \text{otherwise}.
   \end{cases}
\end{aligned}
\end{equation}
Here, $R_{uv} > A_{uv}$, ensuring $G_{uv}(0) < 0$ and $G_{vu}(0) < 0$. Short-range repulsion distances are equal ($d_{uv}^r = d_{vu}^r$), and the long-range attraction for $G_{uv}$ is set to $d_{uv}^a = 3\, d_{uv}^r$. The effective cut-off distance is $d_{uv}^e = d_{vu}^e = 5\, d_{uv}^r$, beyond which no interaction occurs.

\section{Pollutants lead to pigmentation changes}

As pigment patterns in various fish mutants largely result from specific pigment cell–cell interactions, alterations or loss of pigmentation indicate that exposure to environmental pollutants may disrupt these fundamental cellular mechanisms. Motivated by this, we investigate the effects of pollutants on each pigment cell type individually, aiming to identify which interactions are most sensitive to chemical exposure.

\paragraph{Pollutant-induced hyper- and hypopigmentation}

Exposure to certain environmental contaminants can alter fish pigmentation, leading to either hypopigmentation or hyperpigmentation. Some compounds have been reported to stimulate pigmentation and produce darker patterns. For instance, fisetin \cite{molagoda2020gsk}, flumequine \cite{karunarathne2019flumequine}, and rice bran ash mineral extract \cite{kim2019rice} enhance melanogenesis by promoting melanin biosynthesis and transport. Conversely, several contaminants are associated with hypopigmentation. Carbendazim and cadmium, for example, reduce pigment deposition \cite{Schmidt2016,zhang2015mechanisms}. Exposure to N-(1,3-dimethylbutyl)-N$'$-phenyl\emph{p}-phenylenediamine (6PPD) significantly decreases melanin in larval zebrafish \cite{fang20256ppd}. Similarly, the disinfection by-product 3,5-di-\emph{L}-iodotyrosylalanine (DIYA), although not acutely toxic to embryonic zebrafish, significantly inhibits pigmentation in melanophores and xanthophores of the head, trunk, and tail at higher concentrations \cite{peng2020evaluation}.
\begin{figure}[ht!]
    \centering
    \subfigure[]{\includegraphics[width=0.45\textwidth]{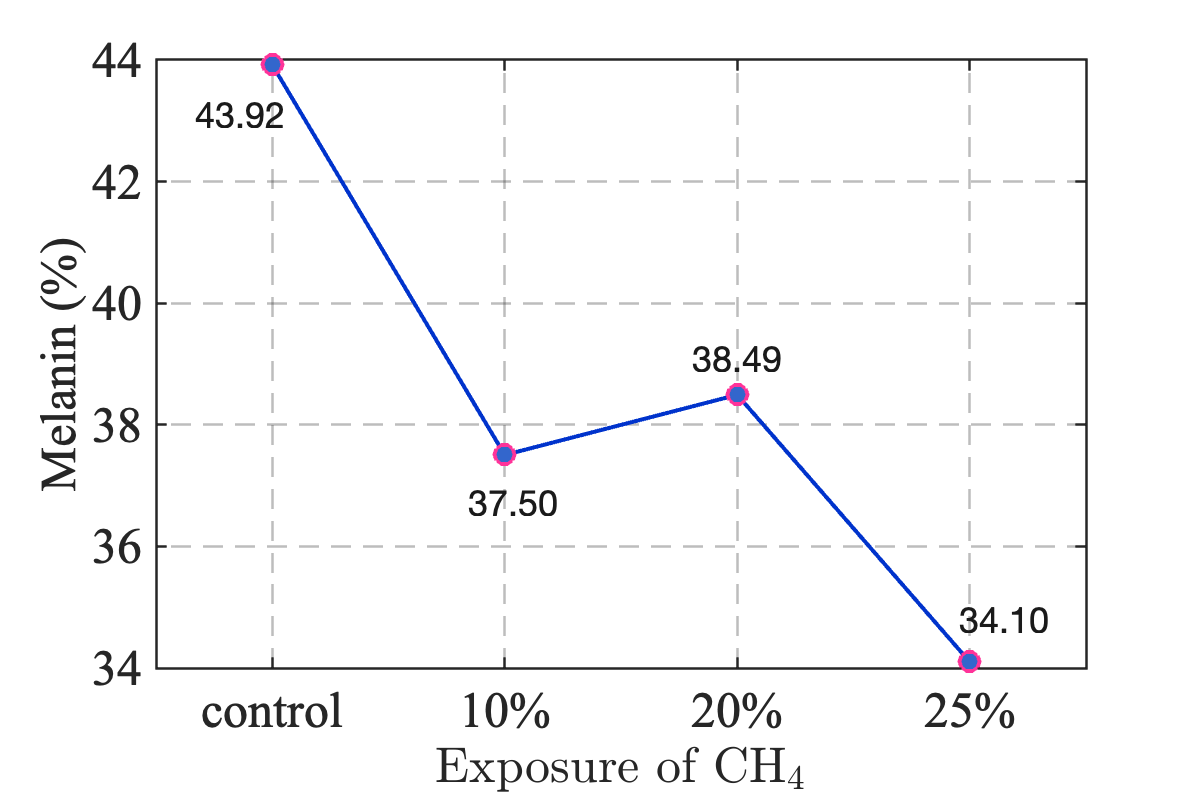}}
    \subfigure[]{\includegraphics[width=0.48\textwidth]{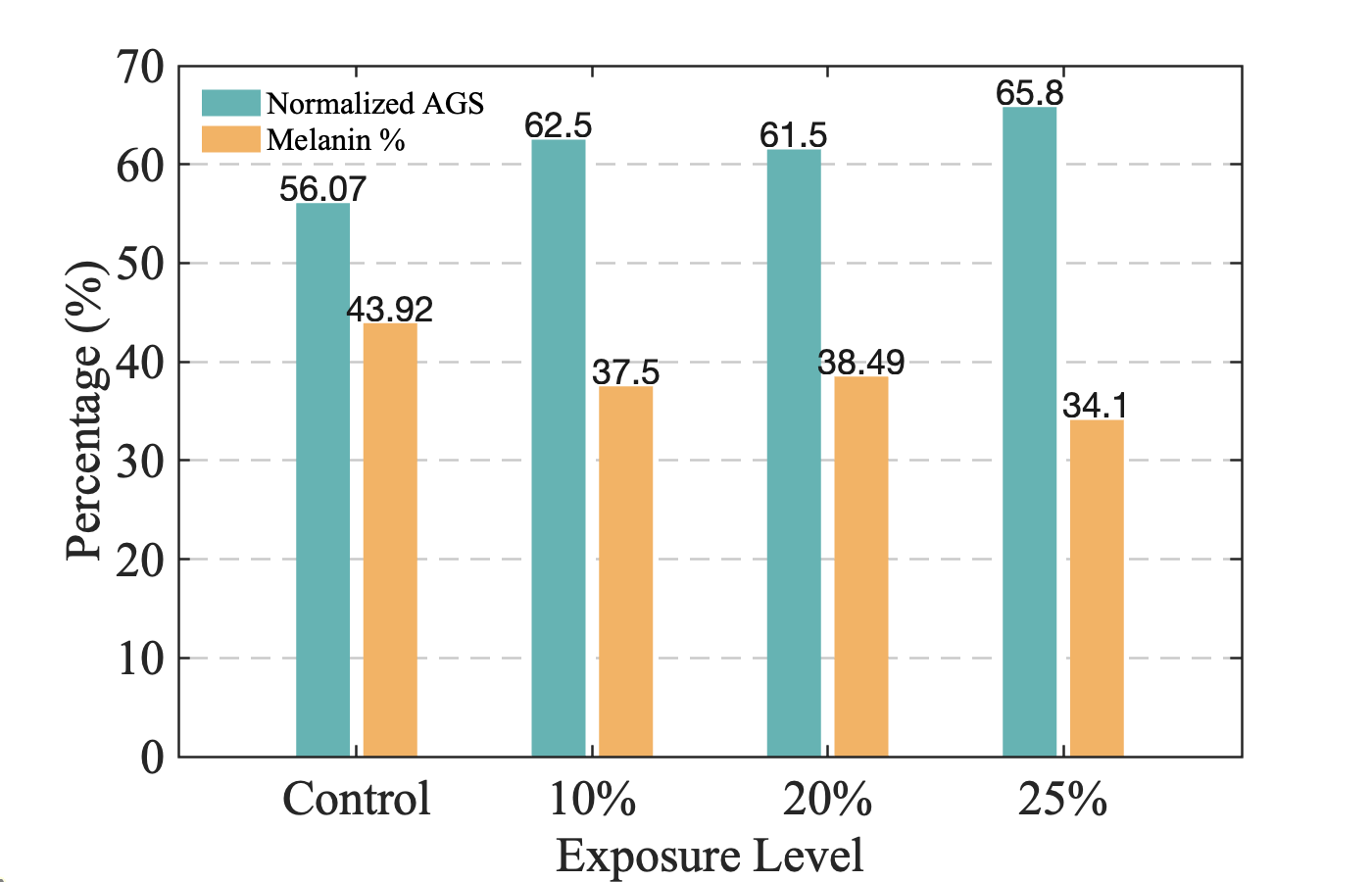}}
    {\includegraphics[width=0.2\textwidth, angle=90]{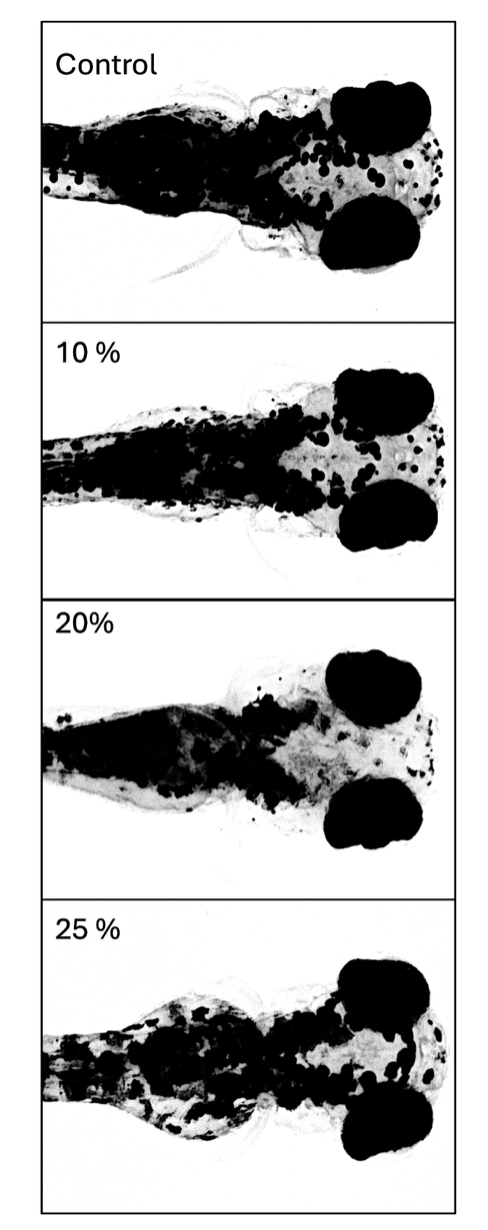}}
\caption{Inhibition of melanin pigmentation in zebrafish larvae at 4 dpf exposed to methane at different concentrations (6-24 fish samples considered for each concentration)  
(a) Melanin percentage in the head-dorsal region at different exposure levels for;  
(b) Histogram of average grayscale (AGS) and melanin percentage for control and methane-exposed larvae;  
(c) Dorsal view of the head region of a zebrafish sample used in the experiment.}
\label{fig:Melanin_AGS}
\end{figure}

\paragraph{Our experimental results: Methane-induced hypopigmentation}\label{Sec:Motivation}

\begin{figure}[ht!]
    \centering
    \includegraphics[width=0.5\linewidth]{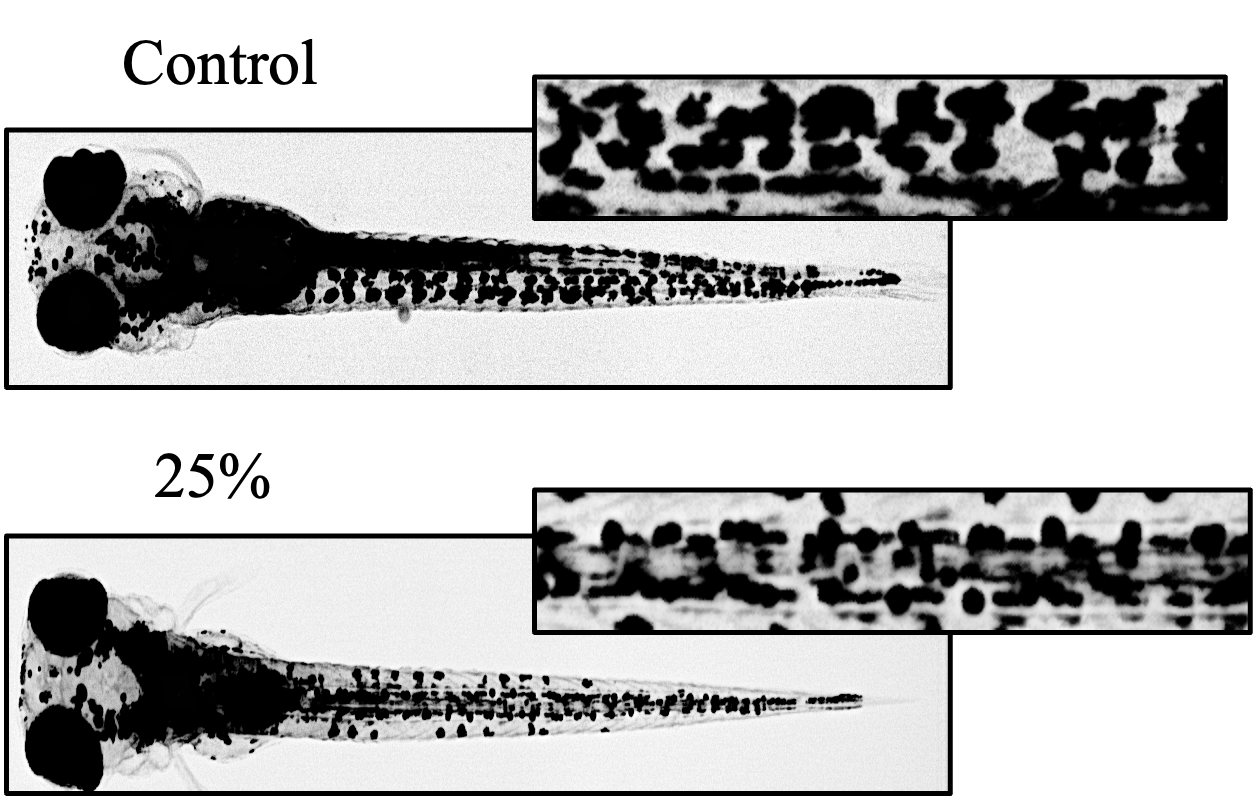}
    \caption{Localized inhibition of pigmentation near the tail region in zebrafish larvae exposed to 25 $\%\, \mathrm{CH}_4$ at 4 dpf compared to control.}
    \label{fig:Pigment_tail}
\end{figure}

In our experiment, zebrafish larvae were exposed to methane ($\mathrm{CH}_4$) for four consecutive days at concentrations of $10\%$, $20\%$, and $25\%$ (by gas volume; oxygen concentration was not altered). After 4 days post-fertilization (dpf), larvae were returned to normal water, while a parallel control group was maintained under identical conditions without methane. Each group consisted of 6–8 larvae. Our results (Fig.~\ref{fig:Melanin_AGS}) show progressive hypopigmentation with increasing methane exposure. The pale coloration reflects delayed melanin development, with quantitative analysis indicating an approximately $10\%$ reduction in melanin in larvae exposed to $25\%$ methane. This inhibitory effect is spatially heterogeneous, with the tail showing the most pronounced reduction, nearly $20\%$ greater than the head (Figs.~\ref{fig:Melanin_AGS} and \ref{fig:Pigment_tail}). Representative images and summarized data are provided in Fig.~\ref{fig:Pigment_tail} and Table~\ref{tab:hypopigmentation} of the supplementary document.

\paragraph{Pollutants can influence cell-cell interactions}

Fish pigmentation patterns (e.g., stripes, spots, and patches) are determined by the number, type, and spatial arrangement of pigment cells, including melanophores, xanthophores, and iridophores \cite{patterson2019zebrafish}. A growing body of evidence shows that environmental pollutants can affect organisms by disrupting fundamental cell–cell interactions \cite{moore1985cellular}. Pollutants are often present at low concentrations and as complex mixtures, and they can produce synergistic, antagonistic, or additive effects that interfere with normal developmental processes. Pollutants can damage cells through multiple pathways, including oxidative stress, mitochondrial dysfunction, inflammation, and proteotoxicity \cite{manzano2023unravelling}. For instance, widely distributed chemicals such as microplastics can adhere to cell membranes, blocking signal transport and impairing protein synthesis \cite{liu2019microplastic}. Other contaminants, including resin acids, cadmium, lead, and monochloramine, can disrupt chemical signaling and induce abnormal cell morphologies \cite{nikinmaa1992does}, while BPA interferes with integrin signaling and lipid raft domains essential for proper cell adhesion and communication \cite{izard2025environmental}. Of specific relevance here, a wide array of organic contaminants can interfere with the cell membrane at its surface, or through partitioning into the 3D interior of the lipid bilayer \cite{roberts2003mechanisms}. These interactions, referred to as polar and non-polar narcosis, can alter the function of membrane bound proteins, which may be key to cell-cell interactions. Regardless of route, Pigment cell systems are highly sensitive to chemical disturbances. Experiments in zebrafish demonstrate that when pigment cells in a localized region are ablated, new cells regenerate the pattern; however, their development and survival strongly depend on interactions with surrounding cells \cite{Nakamasu2009}. Adjacent melanophores and xanthophores exhibit mutual repulsion, maintaining the boundaries of pigment patterns \cite{nakamasu2022correspondences}. These findings collectively suggest that pollutants can perturb a wide range of cellular interaction mechanisms, providing a strong biological rationale for studying how altered intercellular communication may lead to abnormal pigment patterns in developing fish.

\begin{figure}[ht!]
    \centering
    \includegraphics[width=10cm,height=8cm]{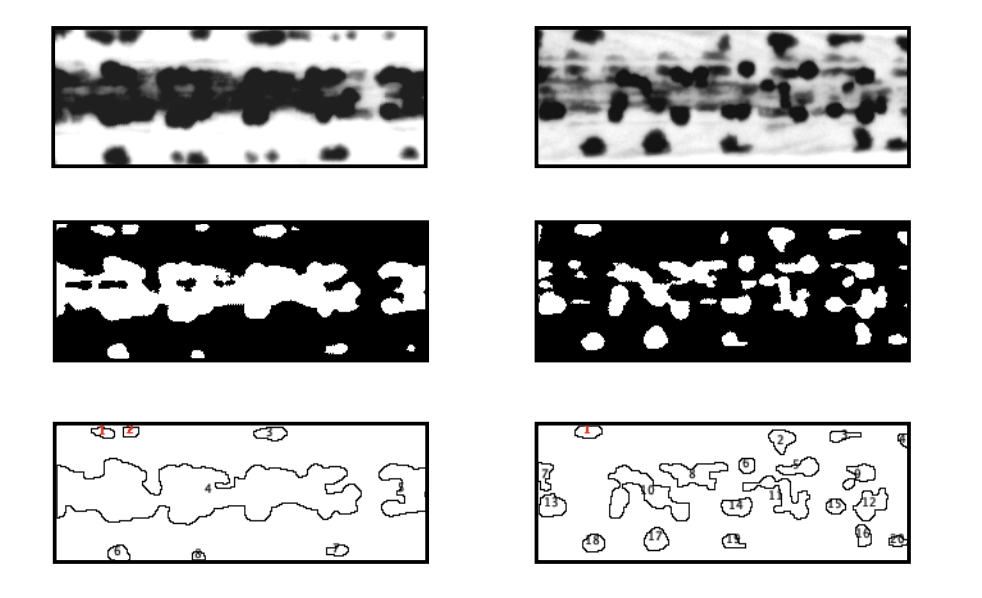}
    \caption{Image-based fragmentation analysis of pigmentation patterns at 4 dpf. The left panel shows continuous stripe development in a control larva, along with the thresholded image and the identified connected pigment regions. The right panel shows a fragmented pigmentation pattern in a larva exposed to $25\%\,\mathrm{CH}_4$, with thresholded images revealing a larger number of spatially disconnected pigment regions.}
    \label{fig:fragmented_exp_data}
\end{figure}

\paragraph{Our experimental results: Methane alters pigment cell--cell interactions}
Quantitative image analysis revealed pronounced pigment fragmentation in 4 dpf zebrafish larvae exposed to $25\%\,\mathrm{CH}_4$. Within the region of interest, control larvae exhibited fewer pigment regions (Count = 8), larger average region size (1080), and greater overall pigment coverage (Area = 29.1\%) (left panel of Fig.~\ref{fig:fragmented_exp_data}). In contrast, methane-exposed larvae showed a substantially higher number of pigment regions (Count = 20), accompanied by a marked reduction in average region size (272) and total pigment coverage (Area = 18.3\%) (right panel of Fig.~\ref{fig:fragmented_exp_data}). Together, these metrics indicate enhanced fragmentation, with pigment cells forming smaller and more spatially disconnected clusters under methane exposure. At 4 dpf, control larvae begin to display a developing continuous stripe in the central body region, whereas methane exposure disrupts this organization, resulting in reduced stripe continuity and impaired early pattern formation. This fragmentation suggests a disruption of normal melanophore aggregation dynamics, which are essential for coherent stripe development. Such effects are consistent with altered adhesive and repulsive pigment cell--cell interactions, as well as impaired contact-dependent signaling, all of which are known to govern pigment pattern formation. Given that methane is relatively unreactive and lipid soluble, it likely acts through membrane-associated mechanisms, such as membrane disruption (narcosis) and/or interference with membrane-embedded proteins.

\paragraph{Pigment cell-cell interactions drive pattern changes}
 Perturbations induced by pollutants or other environmental factors that disrupt this balance will eventually alter adult patterns. For instance, the zebrafish mutant \emph{touchtone}$^{b722}$ (\emph{tct}) lacks neural crest-derived melanophores, and the remaining differentiated melanophores are pale and reduced in size \cite{cornell2004touchtone}. The \emph{nacre} mutant lacks all neural crest-derived melanophores, while \emph{pfeffer} exhibits severely reduced numbers of xanthophores, resulting in the breakdown of stripes into spots. The simultaneous absence of melanophores and xanthophores produces the double mutant \emph{nac;pfe}. In contrast, mutants such as \emph{leopard} and \emph{seurat} display spot-like patterns instead of stripes, whereas \emph{obelix} exhibits broadened stripes \cite{Singh2015}.

\paragraph{Early pigment alterations impact later development}

Many studies examining the effects of pollutants on fish pigmentation focus on larvae, yet early pigment alterations can have long-lasting consequences. Pigment cells specified and patterned during embryonic and early larval stages subsequently undergo differentiation and metamorphosis to form adult pigment patterns \cite{parichy2021evolution}. In zebrafish, neural crest-derived melanophores differentiate during embryogenesis to generate early larval pigment stripes \cite{budi2008embryonic}. This pattern features xanthophores dispersed over the flank, while melanophores and iridophores are positioned along the dorsal and ventral edges, over the yolk, and along a lateral stripe following the horizontal myoseptum \cite{quigley2002pigment}. During metamorphosis, these embryonic melanophores are gradually replaced by newly differentiating metamorphic melanophores, ultimately forming adult stripes \cite{budi2008embryonic}. The early larval pattern is thus remodeled into the adult configuration, consisting of alternating dark stripes of melanophores and iridophores and light stripes of xanthophores and iridophores \cite{quigley2002pigment}. Disruptions during embryogenesis can significantly impact adult pattern formation \cite{wang2022crispr}, as proper stripe development depends on coordinated pigment cell lineage specification, differentiation, specific cellular interactions, and morphogenetic behaviors \cite{patterson2019zebrafish}. On the other hand, even when early larval pigmentation appears normal, disruptions to embryonic signaling pathways may lead to defects in adult patterning. Mutants that display normal embryonic pigmentation but defective adult patterns reveal critical factors for establishing, maintaining, or recruiting metamorphic melanophore precursors. For example, the zebrafish mutant \emph{picasso} lacks most metamorphic melanophores due to mutations in the ErbB gene \emph{erbb3b}, which encodes an EGFR-like receptor tyrosine kinase. This highlights a critical embryonic critical period for ErbB signaling in promoting later pigment pattern metamorphosis, despite normal early melanophore patterns \cite{budi2008embryonic}.

\paragraph{Explanation for pollutant-induced pattern changes}

From a developmental perspective, environmental pollutants may either generate genetic mutations or act as exogenous perturbations that interfere with the cellular processes responsible for normal pigment pattern formation \cite{cavalieri2017environmental,Singh2015,Schmidt2016,zhang2015mechanisms,fang20256ppd}. A plausible mechanism underlying pollutant-induced pigmentation changes is that contaminants entering the organism disrupt developmental signaling pathways, oxidative balance, or cellular metabolism \cite{fairbairn2012polycyclic,wormley2004environmental}. Such disruptions can impair pigment cell specification, migration, survival, or differentiation, thereby altering pigment cell--cell interactions during critical developmental windows. Adult stripe and spot patterns emerge from a finely balanced interplay of adhesive and repulsive interactions among melanophores, xanthophores, and iridophores \cite{patterson2019zebrafish,nakamasu2009interactions}. Pollutant-induced perturbations to this balance can therefore translate into observable changes in pigmentation patterns. Even modest changes in signaling strength, motility, or survival of a single pigment cell type may propagate through intercellular interactions, ultimately leading to large-scale reorganization of the final pigment pattern. This perspective provides a mechanistic link between environmental contamination, developmental toxicity, and the emergence of abnormal pigmentation patterns in adult fish.

\section{Results}


To investigate the effects of pollutants on fish pigmentation patterns, we incorporate pollutant influence into the model. Assuming that pollutants are approximately uniform within a local region, we represent the pollutant level as a constant pollution level $T_0$ with small random fluctuations: 
\[
T(x,y) = T_0 + \varepsilon \,\xi(x,y),
\] 
where $\xi(x,y)$ is a uniform random variable on $[0,1]$, and $\varepsilon$ is a small perturbation.

Different pollutants can have diverse effects on pigmentation, including reduced pigmentation, hyperpigmentation, or irregular patterns \cite{zhou2022beta,hayazaki2021zebrafish}. Since fish stripe and spot patterns emerge from a finely balanced interplay of adhesive and repulsive interactions among melanophores, xanthophores, and iridophores \cite{nakamasu2009interactions}, these observations suggest that pollutants can perturb this balance in multiple ways. Within a reaction–advection–diffusion framework, such exposure can be represented as changes in interaction kernel strengths. We therefore examine representative cases of homotypic and heterotypic interactions under pollutant exposure to understand how alterations in pigment cell interactions lead to diverse pigmentation patterns. While experimental studies provide valuable observations, mathematical modeling enables a systematic exploration of how specific changes in cell–cell interactions propagate to influence overall pattern formation. This approach also allows us to investigate the effects of varying pollutant levels, timing, or modes of action, identify interactions most sensitive to disruption, and test hypothetical scenarios that are difficult to study experimentally.



\subsection{Pollutant-induced changes in homotypic and heterotypic cell interactions}
\label{sec:Pollutant-induced changes in homotypic and heterotypic cell interactions}
\begin{figure}
    \centering
    \includegraphics[width=0.65\linewidth]{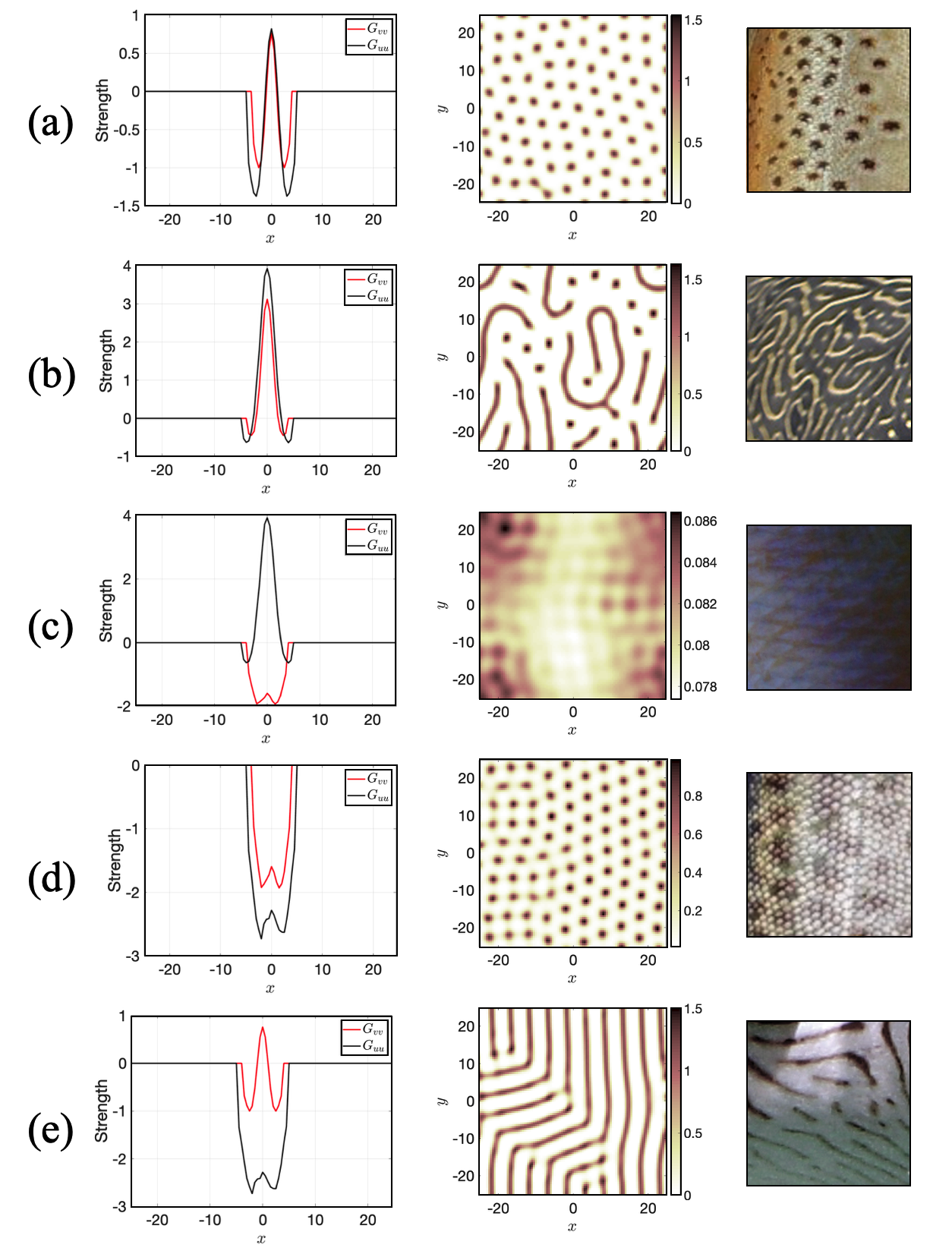}
    \caption{
Comparison of simulated pigment patterns at $t = 300$ for different cases in which the pollutant modifies cell–cell interactions. 
Panels (a–e) correspond to the control (no pollutant), HM1, HM2, HM3, and HM4 scenarios, respectively. 
The first column shows the potential (kernel) functions that describe the attractive and repulsive interaction strengths and ranges for each case. 
The second column displays the numerically generated patterns, and the third column shows representative mutant fish skin exhibiting qualitatively similar pattern types (images adapted and cropped from Wikipedia). 
Parameters: $R_{vv} = 3$, $A_{uu} = 5$, $A_{vv} = 4$, $d_{uu} = 5$, $d_{vv} = 3$.  
The polluted field is given by $T(i,j) = 0.5 + 0.1\,\xi_1(i,j)$.  
Initial conditions: $u_0(i,j) = 1 + 0.01\,\xi_2(i,j)$ and $v_0(i,j) = 1 + 0.02\,\xi_3(i,j)$, where $\xi_k(i,j) \sim \mathcal{N}(0,1)$.} 
\label{fig:homotypic-kernels-patterns}
\end{figure}
Heterotypic interactions primarily determine pigment cell morphology, while homotypic interactions regulate cell proliferation and dispersal \cite{walderich2016homotypic}. Together, these interactions establish the spatial arrangement of pigment cells that underlies the characteristic color patterns in fish. Here, we examine how pollutant-induced disruptions in cell–cell signaling affect both homotypic and heterotypic interactions. We first consider several representative cases of homotypic interactions under pollutant exposure, focusing on interactions between cells of the same type.

\textbf{(HM1)} The pollutant reduces repulsion between cells while enhancing short-range adhesion until reaching a saturating threshold. We model this as
\begin{equation*}
    \begin{aligned}
       & R_{uu}(T(x,y)) = R_{uu}\left( \frac{1}{1+T(x,y)}\right),\quad
         R_{vv}(T(x,y)) = R_{vv}\left( \frac{1}{1+T(x,y)}\right),\\
       & A_{uu}(T(x,y)) = A_{uu}\left(1+ \frac{T(x,y)}{1+T(x,y)}\right),\quad
         A_{vv}(T(x,y)) = A_{vv}\left(1+ \frac{T(x,y)}{1+T(x,y)}\right).
    \end{aligned}
\end{equation*}
The saturating functional form is motivated by the dose–response curve of hypopigmentation with exposure concentration (see supplementary information Fig.~\ref{fig:Hypopigment_methane}). Here, the net attractive potential among $u-u$ cells is greater than that of $v-v$ cells. As shown in Fig.~\ref{fig:homotypic-kernels-patterns}(b), spot patterns partially break down, producing a mixture of spots and stripes.

\textbf{(HM2)} The pollutant increases melanophore attraction while decreasing their repulsion, with the opposite effect for xanthophores (increased repulsion and reduced attraction). This combination of $R_{uu},\,A_{uu},\,R_{uv},\,A_{uv}$ ensures strong mutual attraction among melanophores and strong repulsion among xanthophores. Despite enhanced melanophore attraction, black spot patterns disappear, indicating that xanthophore repulsion is critical for maintaining black pigmentation (cf.~Fig.~\ref{fig:homotypic-kernels-patterns}(c)).

\textbf{(HM3)} The pollutant increases net repulsion for both xanthophores and melanophores while reducing attraction. Although a spot pattern similar to the control persists, the maximum melanophore density ($u$) is significantly lower, and the overall coloration is lighter, indicating hypopigmentation (cf.~Fig.~\ref{fig:homotypic-kernels-patterns}(d)).

\textbf{(HM4)} Without influencing xanthophore interactions, the pollutant only increases repulsion and decreases attraction among melanophores as contaminant concentration rises:
\begin{equation*}
    R_{uu}(T(x,y)) = R_{uu}\left(1 + \frac{T(x,y)}{1+T(x,y)}\right),\quad
    A_{uu}(T(x,y)) = A_{uu}\left(\frac{1}{1+T(x,y)}\right).
\end{equation*}
In this case, melanophores experience a net repulsive potential ($G_{uu}<0$), causing the spot pattern from the control to break down into irregular stripe patterns (upper panels of Fig.~\ref{fig:homotypic-kernels-patterns}).

As can be seen in Figure \ref{fig:homotypic-kernels-patterns}, different pollutant-induced changes in cell interactions can generate diverse “mutant-like” body patterns. From maximum melanophore densities, cases HM2 and HM3 produce hypopigmentation,  while a purely repulsive potential among xanthophores can induce hypopigmentation without disrupting the overall pattern structure. These results demonstrate how pollutants modulate cell–cell interactions to drive variations in fish pigmentation.

\begin{figure}[ht!]
\centering
\scriptsize
\setlength{\tabcolsep}{-8pt}
\renewcommand{\arraystretch}{0}
\captionsetup[subfigure]{labelformat=simple, labelsep=none, skip=-4pt}

\subfigure[No pollutant ]{\includegraphics[width=0.2\textwidth, trim=5 5 5 5, clip]{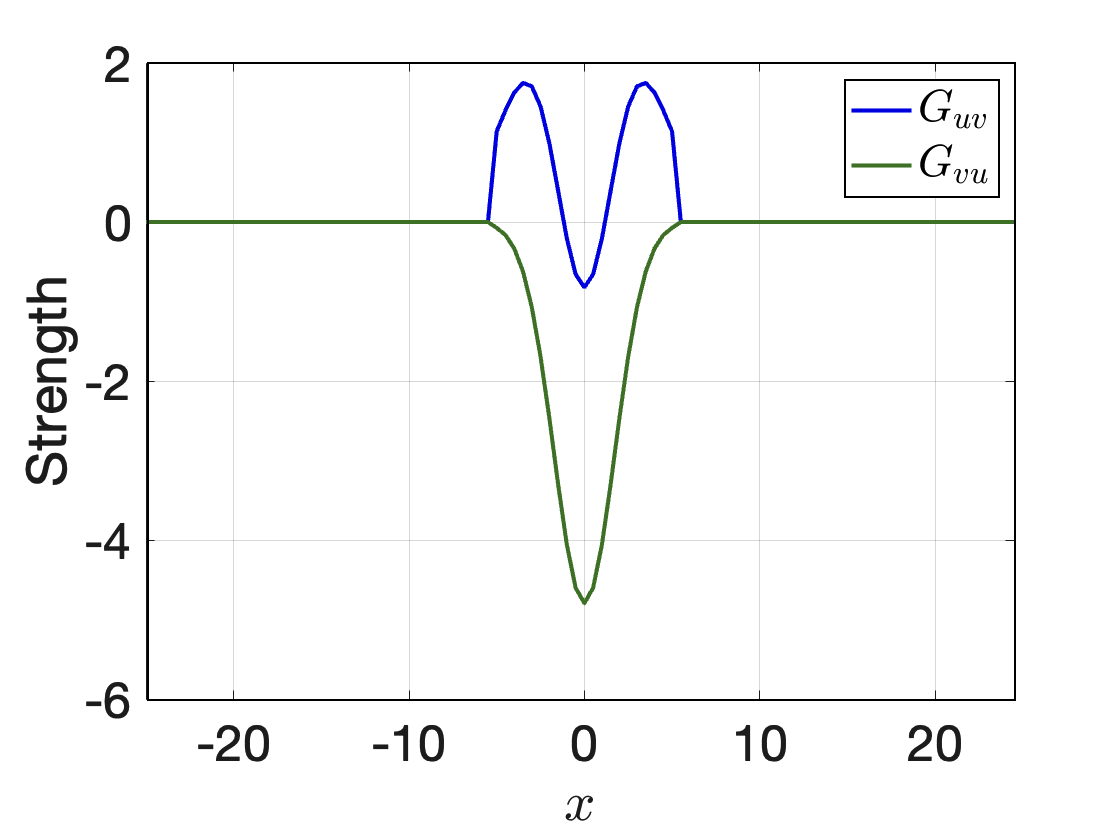}}
\subfigure[HT1 kernel]{\includegraphics[width=0.2\textwidth, trim=5 5 5 5, clip]{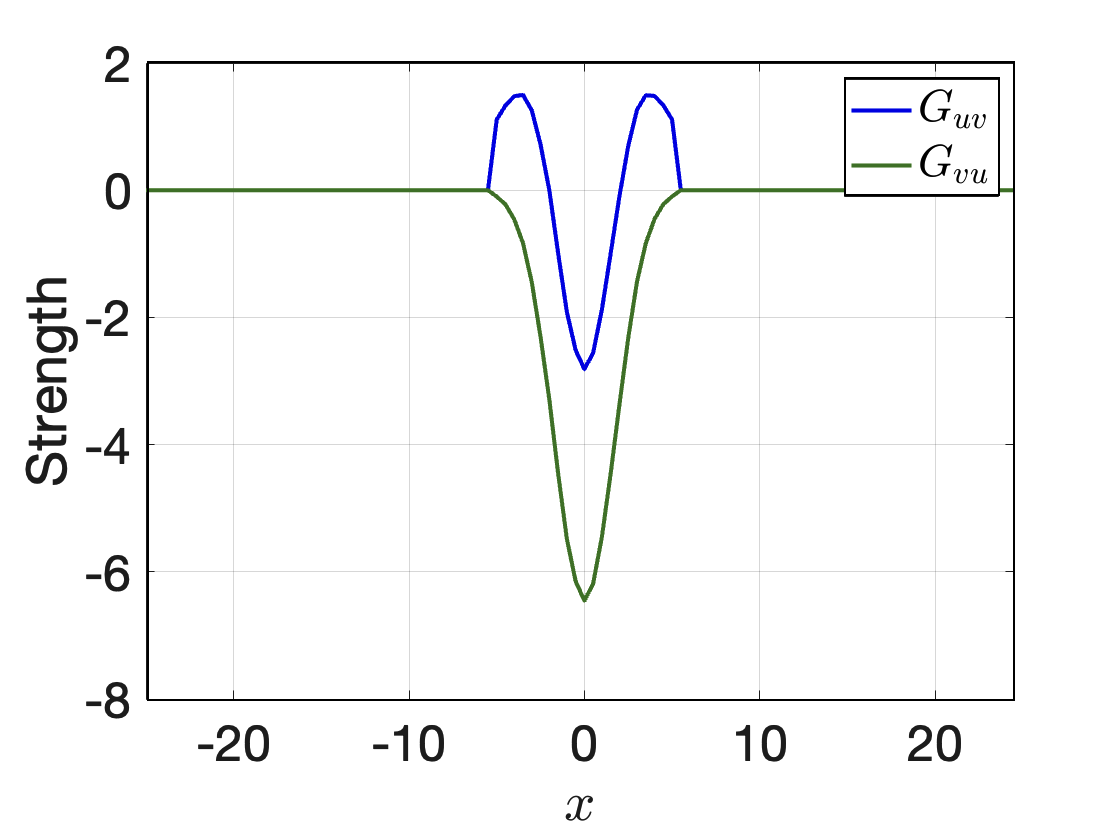}}\hspace{-10pt}
\subfigure[HT2 kernel]{\includegraphics[width=0.2\textwidth, trim=5 5 5 5, clip]{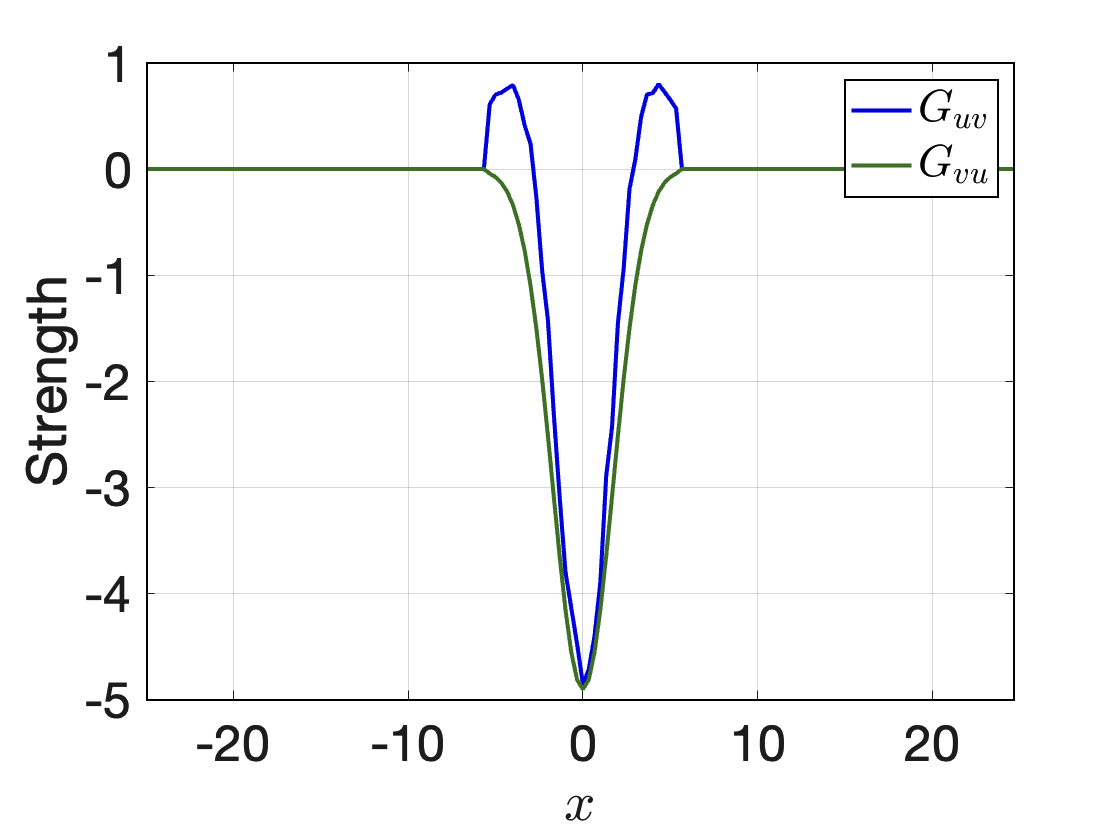}}\hspace{-10pt}
\subfigure[HT3 kernel]{\includegraphics[width=0.2\textwidth, trim=5 5 5 5, clip]{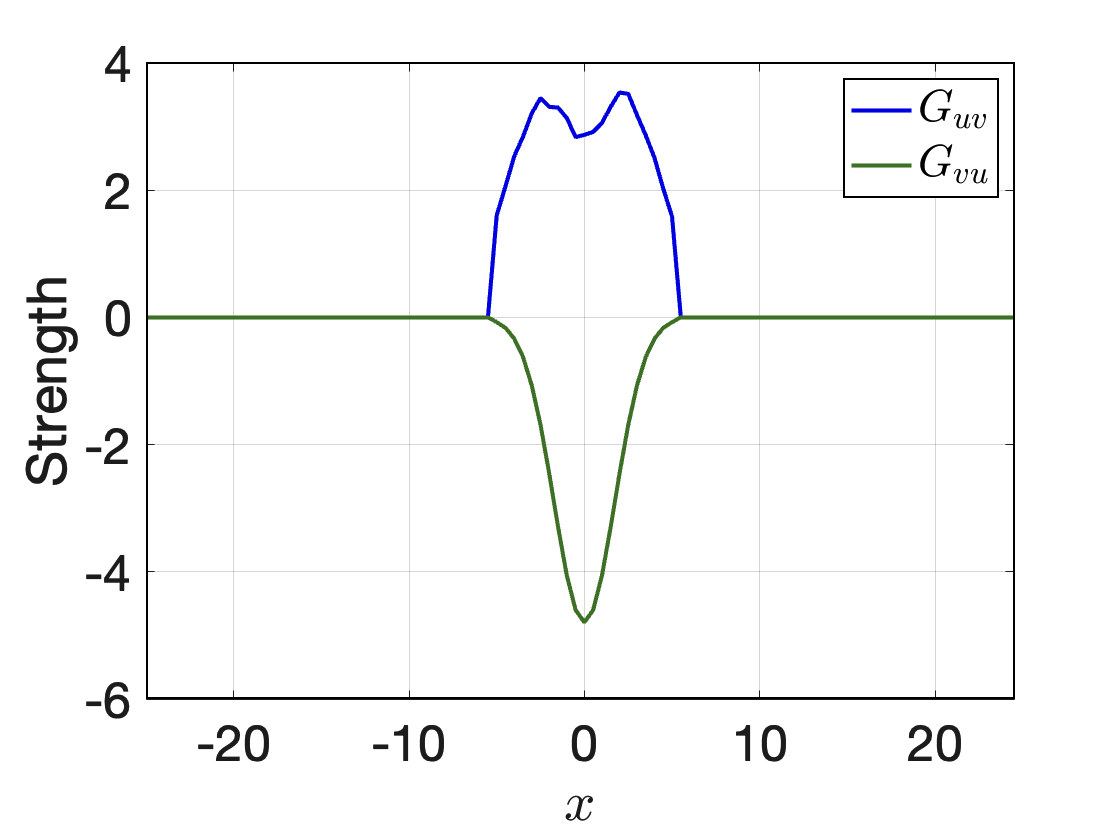}}\hspace{-10pt}
\subfigure[HT4 kernel]{\includegraphics[width=0.2\textwidth, trim=5 5 5 5, clip]{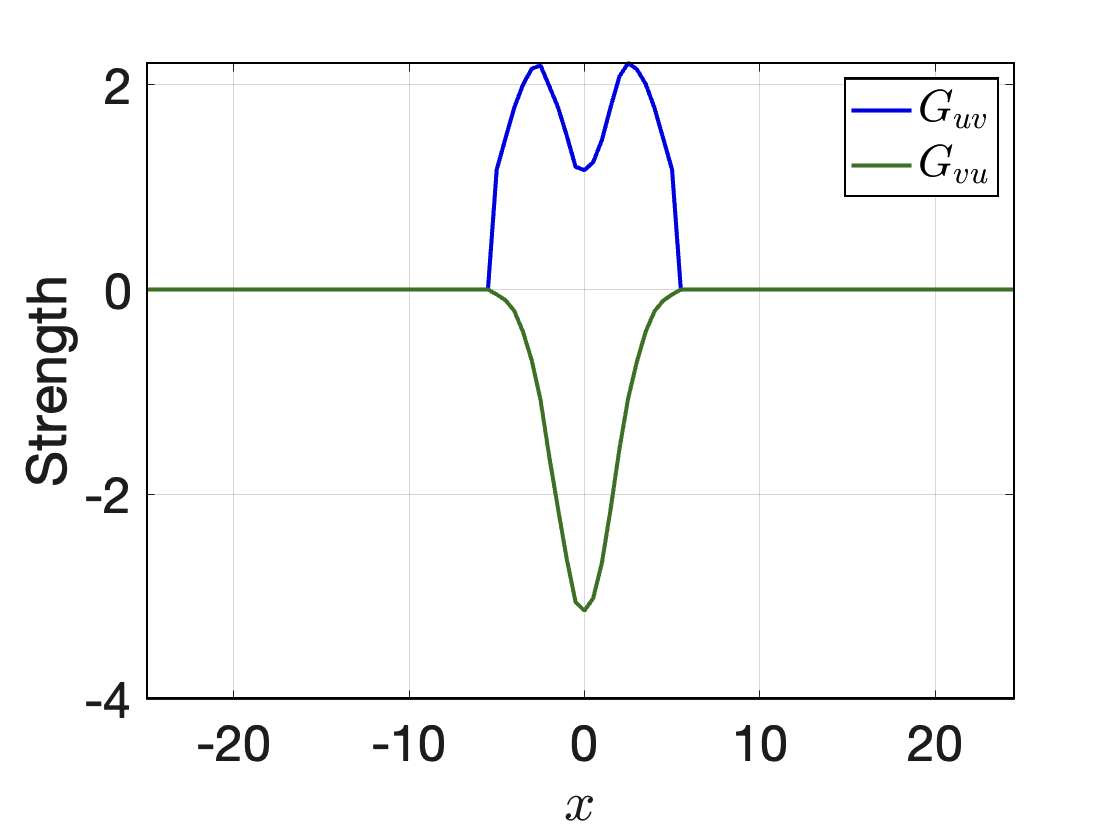}}\hspace{-10pt}

\subfigure[No pollutant ]{\includegraphics[width=0.2\textwidth, trim=5 5 5 5, clip]{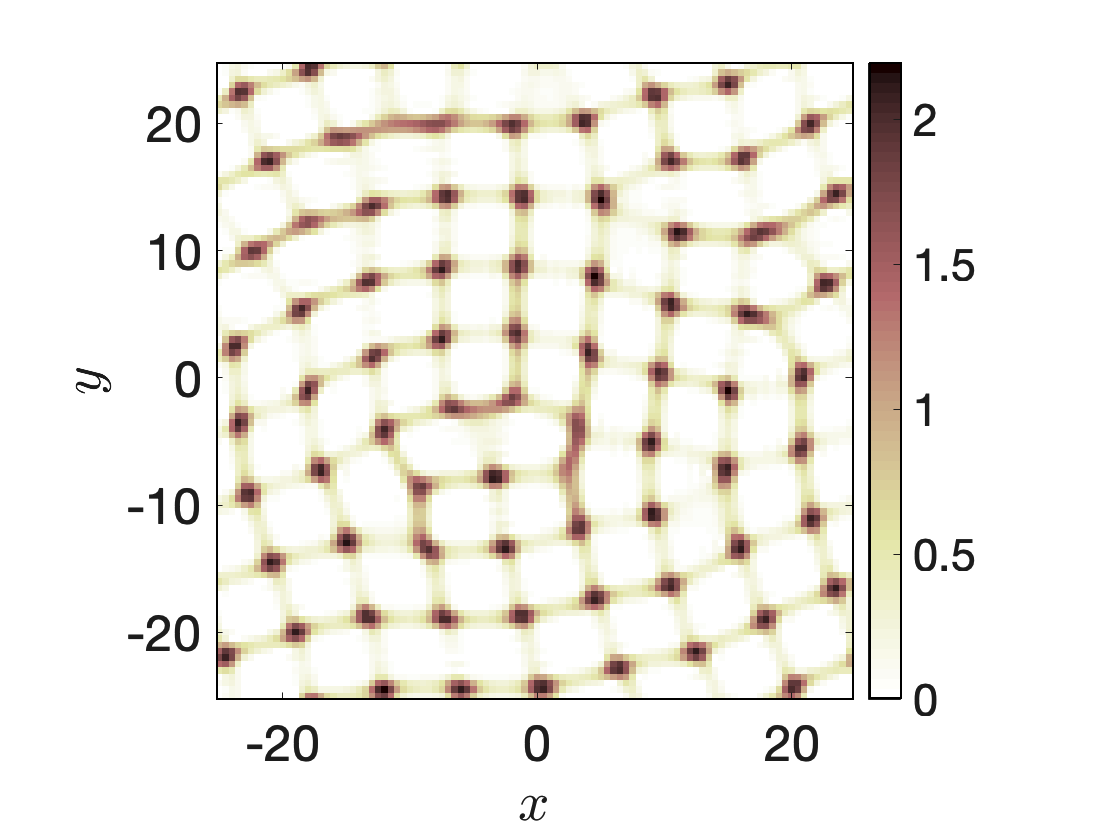}}
\subfigure[HT1 ]{\includegraphics[width=0.2\textwidth, trim=5 5 5 5, clip]{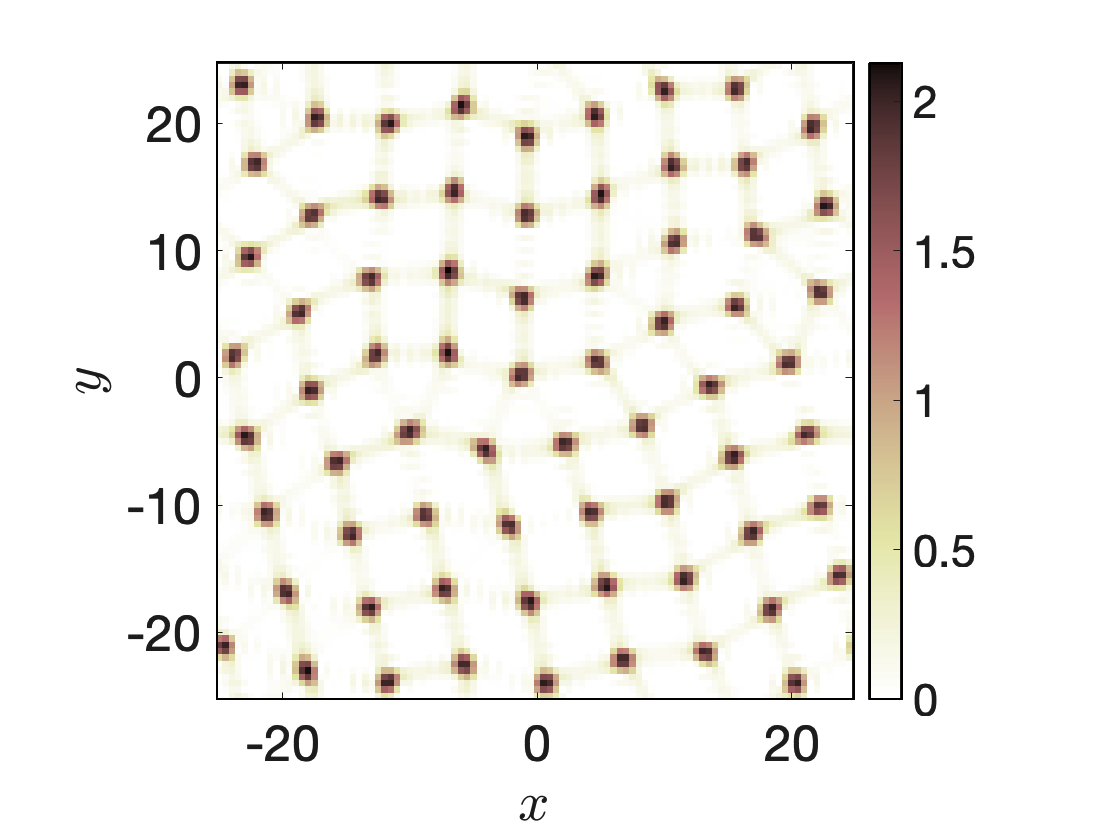}}\hspace{-10pt}
\subfigure[HT2 ]{\includegraphics[width=0.2\textwidth, trim=5 5 5 5, clip]{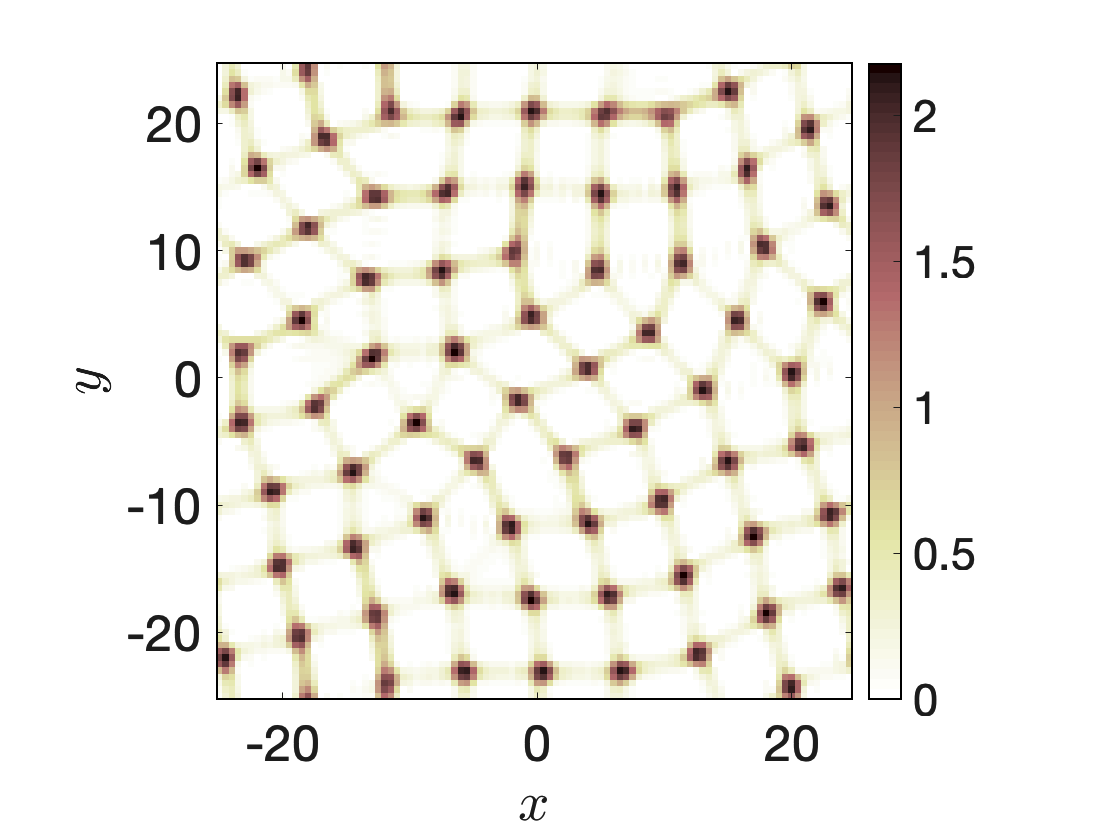}}\hspace{-10pt}
\subfigure[HT3 ]{\includegraphics[width=0.2\textwidth, trim=5 5 5 5, clip]{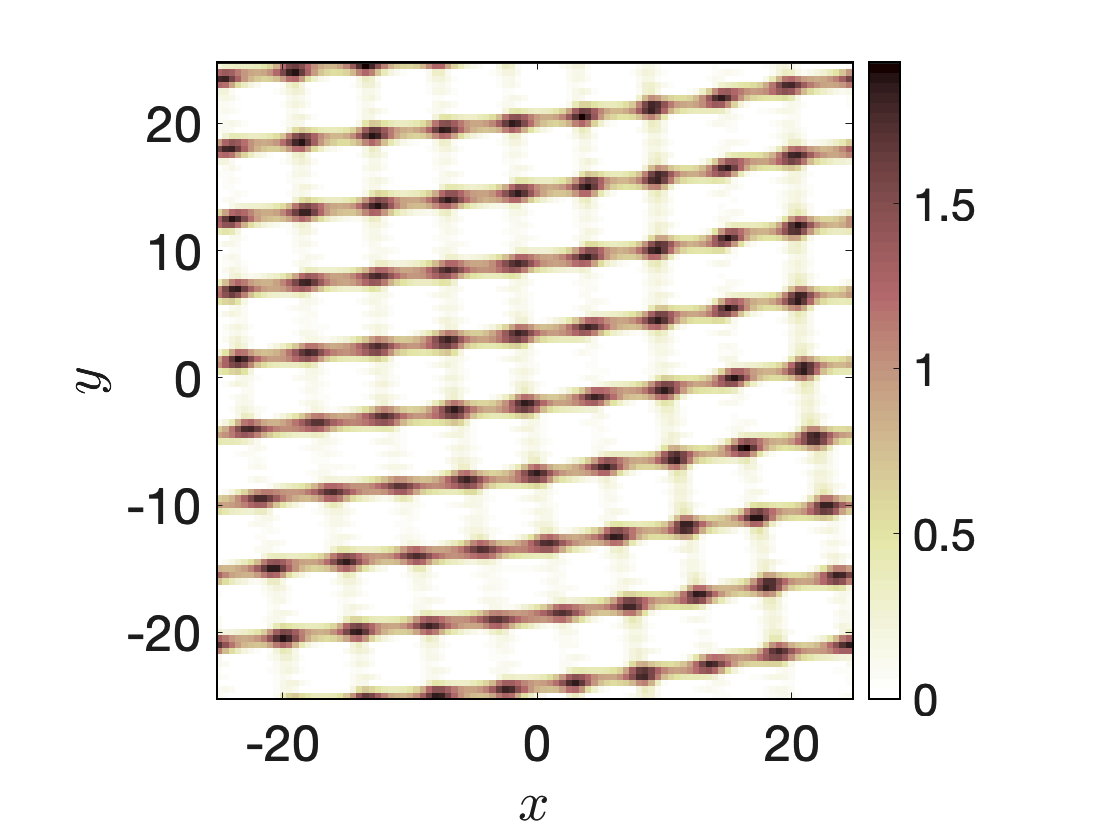}}\hspace{-10pt}
\subfigure[HT4 ]{\includegraphics[width=0.2\textwidth, trim=5 5 5 5, clip]{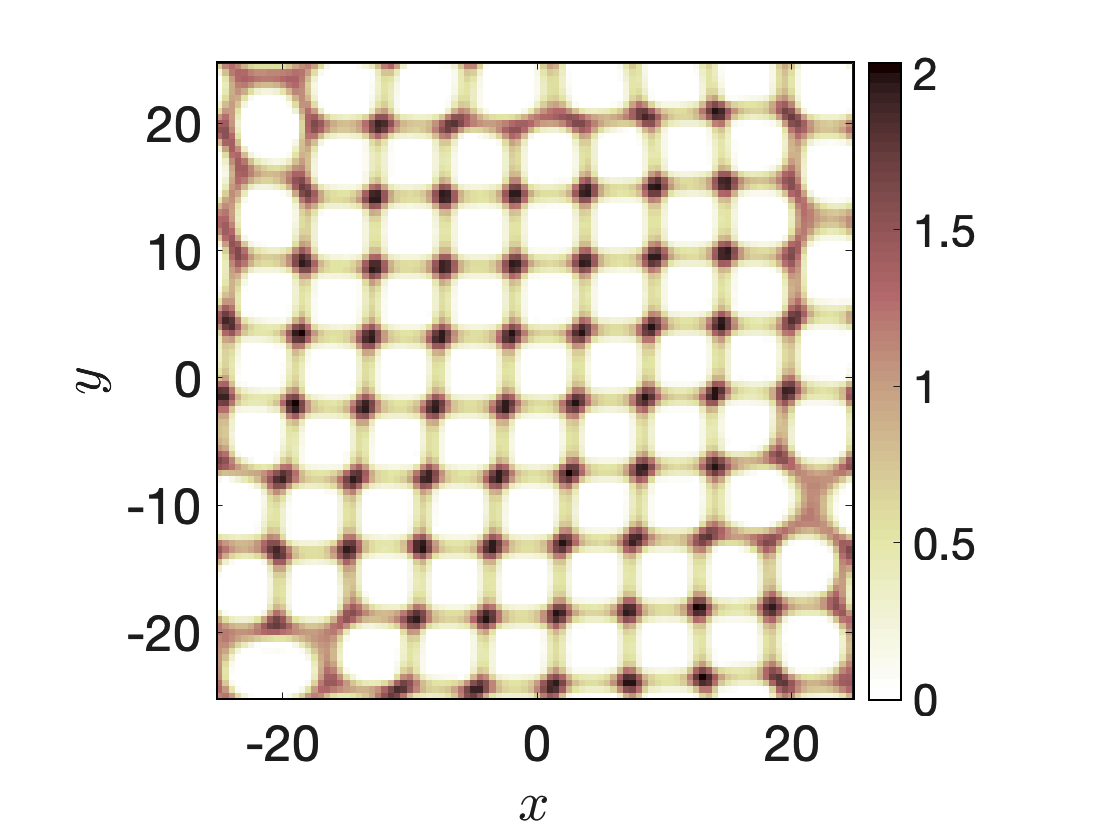}}\hspace{-10pt}

\vspace{-8pt}
\caption{Alteration of patterns considering the heterotypic interaction between the black and yellow cells at $t=300$. 
(a-e) The change in the potential functions (kernel) $G_{uv}$ and $G_{vu}$ due to incorporation of external pollutant; (f-j) The corresponding pattern formation; 
The homotypic parameters $R_{vv}$, $A_{uu}$, $A_{vv}$, $d_{uu}$, $d_{vv}$ are kept fixed as in Fig.~\ref{fig:homotypic-kernels-patterns}, and the heterotypic attractive and repulsion strengths are $R_{uv}=6,\,A_{uv}=5,\,R_{vu}=5$ and $A_{vu}=0,$ and the interaction range $d_{uv}=6.$
The polluted field $T(i,j)= 0.5 + 0.1\,\xi_1(i,j)$.  The initial conditions: $u_0(i,j) = 1 + 0.01\,\xi_2(i,j)$ and $v_0(i,j) = 1 + 0.02\,\xi_3(i,j),$ and $\xi_k(i,j) \sim \mathcal{N}(0,1)$; }

\label{fig:heterotypic-kernels-patterns}
\end{figure}

We next investigate the effects of pollutants on heterotypic interactions. Specifically, we consider several representative scenarios in which pollutant exposure alters the attractive and repulsive forces governing heterotypic interaction potentials. The resulting interaction kernels and corresponding pattern outcomes are shown in Fig.~\ref{fig:heterotypic-kernels-patterns}, where distinct potential profiles emerge under different pollutant-induced perturbations.

\textbf{(HT1)} Pollutant exposure enhances the short-range repulsion exerted by melanophores on xanthophores. 

\textbf{(HT2)} The pollutant simultaneously increases the repulsion exerted by melanophores on xanthophores while weakening their long-range attraction, rendering the attractive and repulsive contributions comparable in magnitude. 

\textbf{(HT3)} The pollutant strengthens the long-range attraction of melanophores toward xanthophores while reducing short-range repulsion. 

\textbf{(HT4)} Pollutant exposure reduces both short-range repulsive interactions—melanophore-on-xanthophore and xanthophore-on-melanophore. 

\begin{figure}[ht!]
\centering
\scriptsize
\setlength{\tabcolsep}{-8pt}
\renewcommand{\arraystretch}{0}
\captionsetup[subfigure]{labelformat=simple, labelsep=none, skip=-4pt}
\includegraphics[width=7cm,height=5cm]{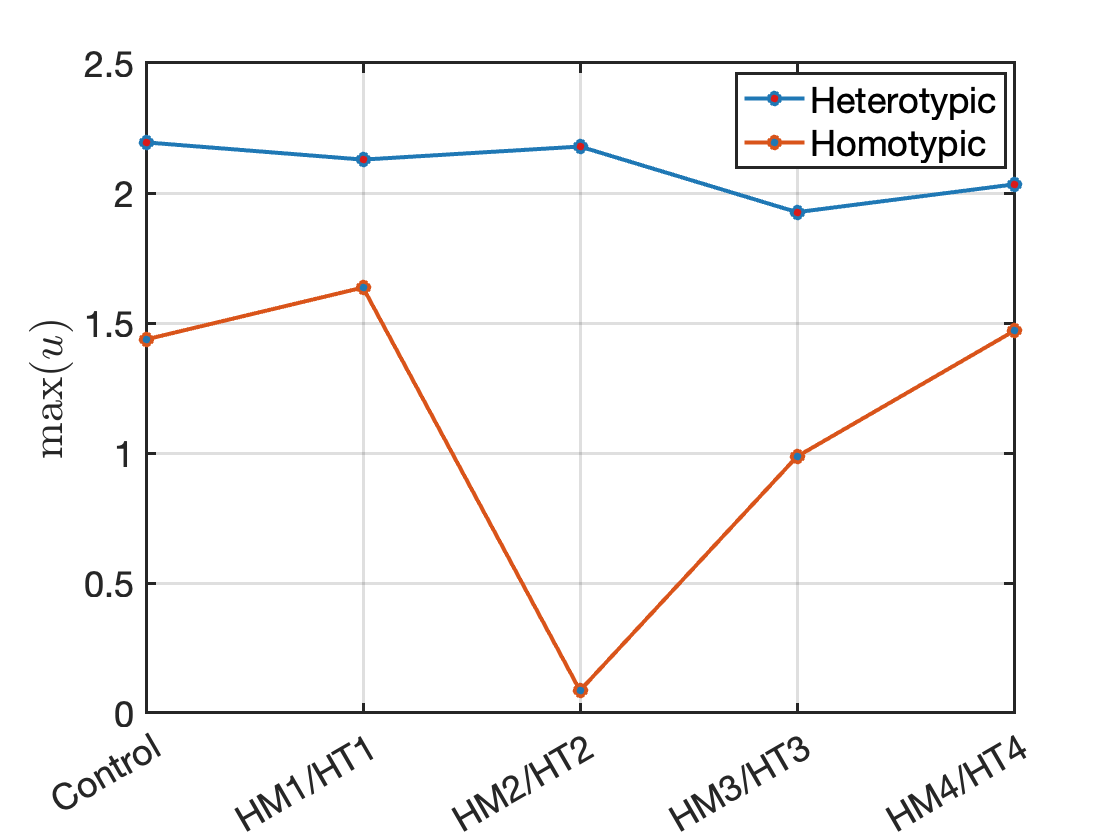}
\caption{The maximum melanophore density for the patterns obtained in homotypic and heterotypic interactions.}
\label{fig:pattern_statistics}
\end{figure}

For all four heterotypic cases, although pollutant exposure induces distinct modifications to interaction strengths, the resulting patterns remain largely similar to the pollutant-free control. Among these scenarios, only HT3 leads to noticeable hypopigmentation. Compared with the homotypic cases shown in Fig.~\ref{fig:homotypic-kernels-patterns}, pollutants have a weaker overall effect on melanophore–xanthophore interactions, indicating that homotypic interactions are considerably more sensitive to pollutant-induced perturbations. This is further reflected in the maximum densities of melanophores ($u$) and xanthophores ($v$) at $t = 300$ across the different hypotheses (Fig.~\ref{fig:pattern_statistics}). In homotypic scenarios, the maximum melanophore density varies widely, from approximately 0.1 to 1.7, with changes of up to 90\%, whereas in heterotypic cases it remains within a narrower range of 1.91–2.21, with the maximum change across all cases being no more than 9\% compared with the control.

\subsection{Saturation effects of pollutants on pigmentation}

\begin{figure}[ht!]
\centering
\footnotesize  
\renewcommand{\arraystretch}{1}  

\subfigure[]{
\includegraphics[width=0.45\textwidth, trim=5 0 5 5, clip]{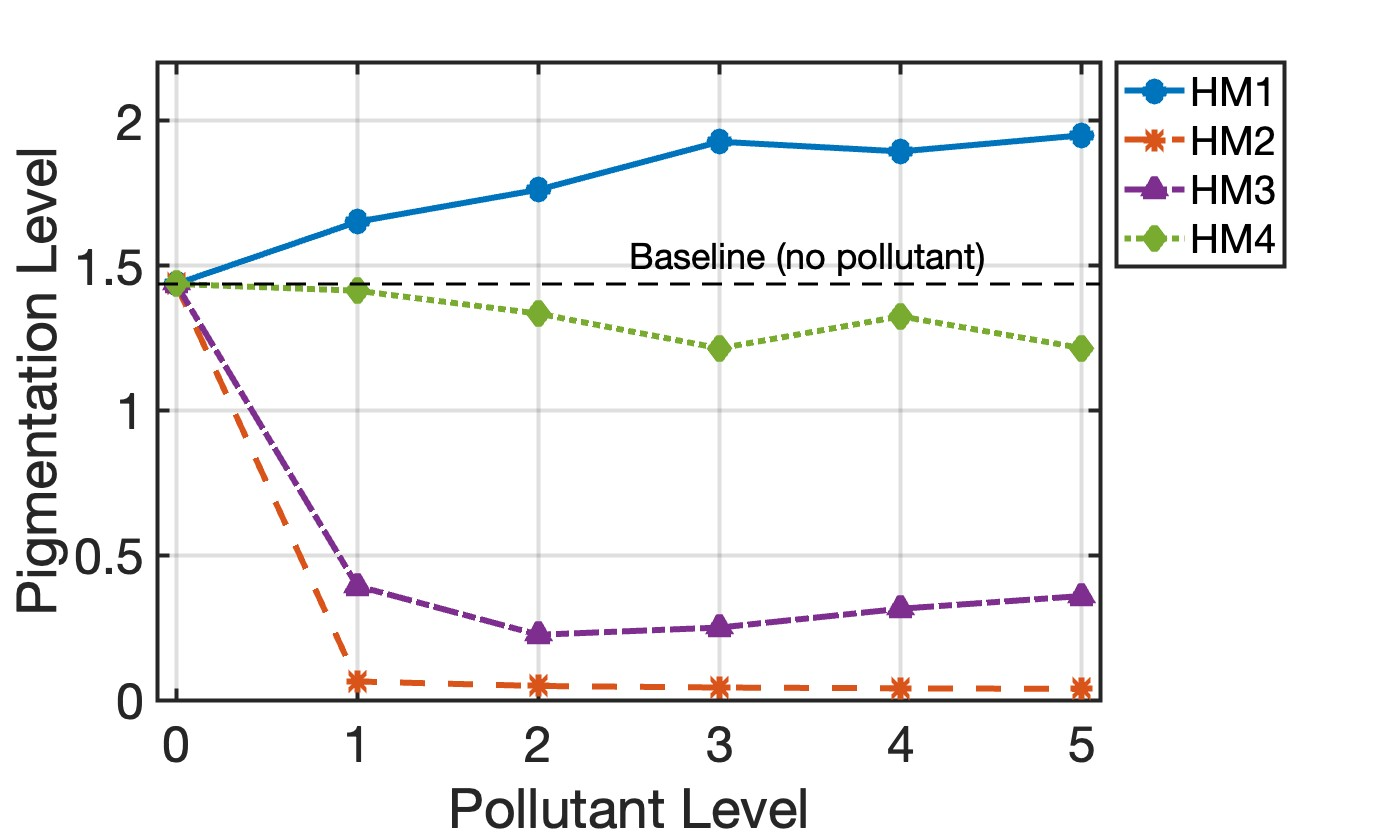}}
\subfigure[]{
\includegraphics[width=0.45\textwidth, trim=5 0 5 5, clip]{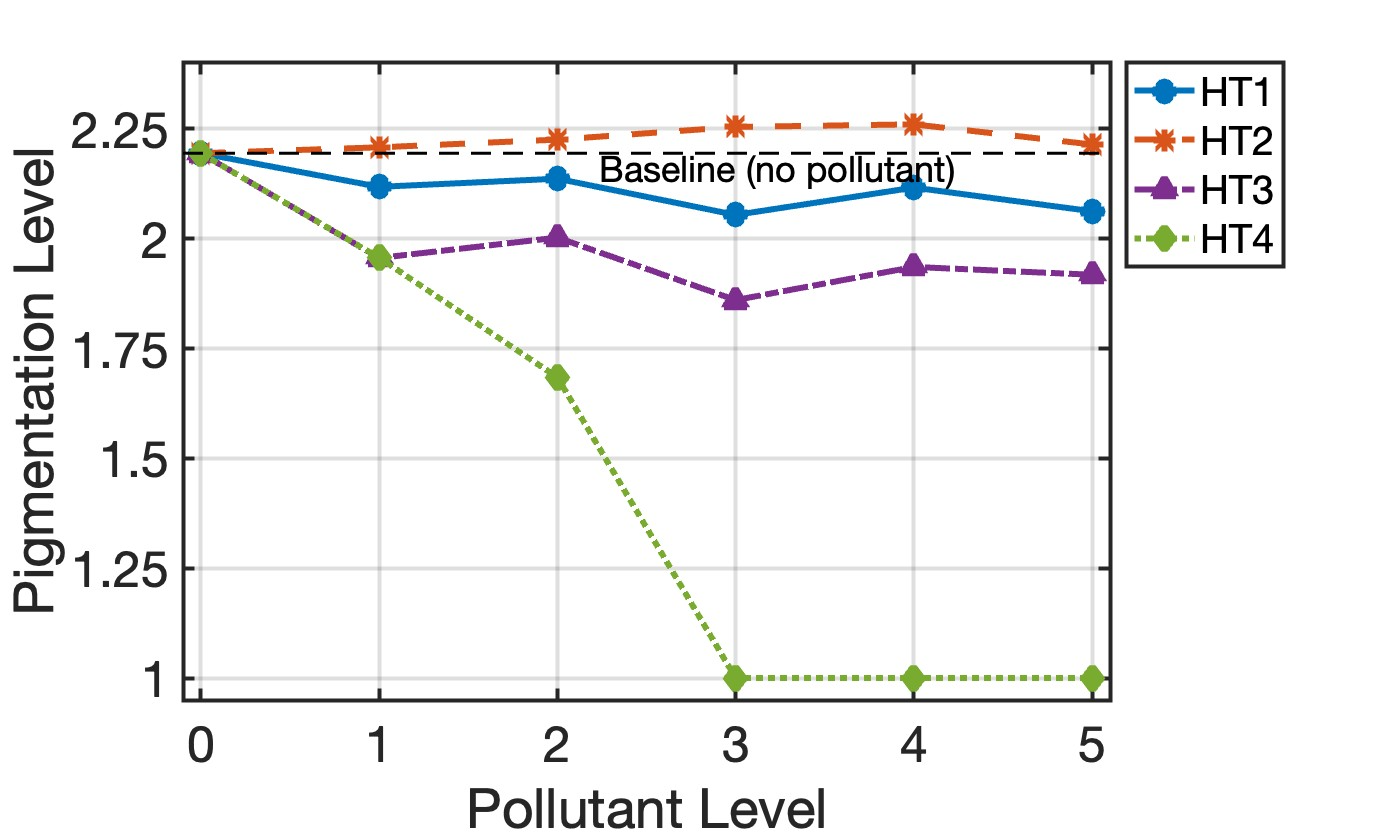}}

\vspace{-5pt}
\caption{(a) Comparison of pigmentation at $t = 300$ for different pollutant  levels under hypotheses HM1--HM4 (homotypic case). 
(b) Comparison of pigmentation at $t = 300$ for different pollutant  levels under hypotheses HT1--HT4 (heterotypic case). 
Parameters: $R_{vv} = 3$, $A_{uu} = 5$, $A_{vv} = 4$, $d_{uu} = 5$, $d_{vv} = 3$. 
The polluted field is defined as $T(i,j) = T_0 \,(1 + 0.1\,\xi_1(i,j))$, with pollutant levels $T_0 = 1, 2, 3, 5$. 
Saturation effects are observed at very high pollutant levels, consistent with the experimental findings reported in \cite{Schmidt2016}. 
Initial conditions are given by $u_0(i,j) = 1 + 0.01\,\xi_2(i,j)$ and $v_0(i,j) = 1 + 0.02\,\xi_3(i,j)$, where $\xi_k(i,j) \sim \mathcal{N}(0,1)$.}
\label{fig:hypopigmentation_summary}
\end{figure}
The different cases of cell interaction changes were tested under varying levels of pollutant exposure, using the control (no-pollutant) case as a baseline. We found that as pollutant levels increase, the qualitative nature of the pigmentation response, whether hypopigmentation or hyperpigmentation, remains unchanged, while the magnitude of the effect generally becomes stronger (cf.~Fig.~\ref{fig:hypopigmentation_summary}). Specifically, in case HM1, increasing pollutant levels ($T_0 = 1,2,\dots,5$) consistently lead to hyperpigmentation. In contrast, HM2, HM3, HM4, as well as HT3 and HT4, consistently exhibit hypopigmentation, with the effect becoming more pronounced as pollutant concentration increases within a certain range (cf.~Fig.~\ref{fig:hypopigmentation_summary}(b)). However, as pollutant levels continue to rise, pigmentation changes, whether increasing or decreasing, eventually reach a saturation point and do not intensify indefinitely. This saturation behavior is consistent with our methane experimental findings.

\subsection{Short-time vs.~continuous exposure to pollutants}

\begin{figure}[ht!]
\centering
\scriptsize

\subfigure[$t=150$]{
    \includegraphics[width=0.31\textwidth]{\detokenize{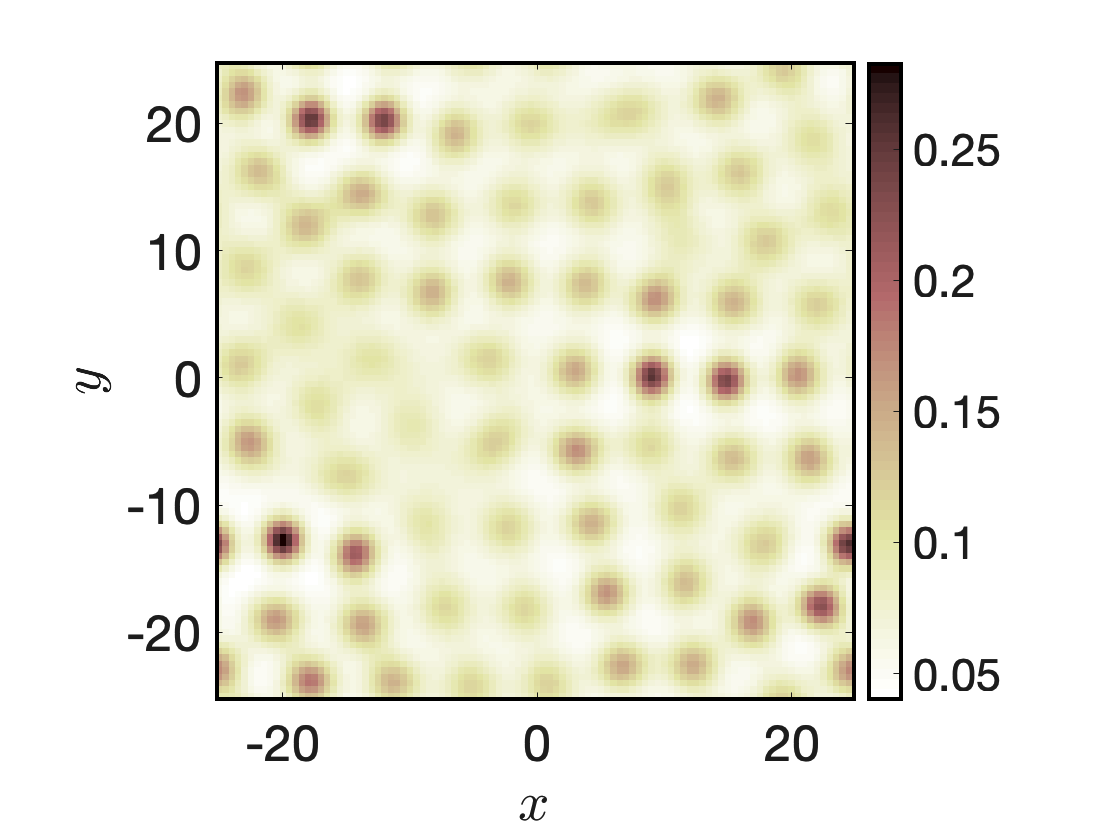}}
}
\hspace{-4pt}
\subfigure[two-phase, $t=300$]{
    \includegraphics[width=0.31\textwidth]{\detokenize{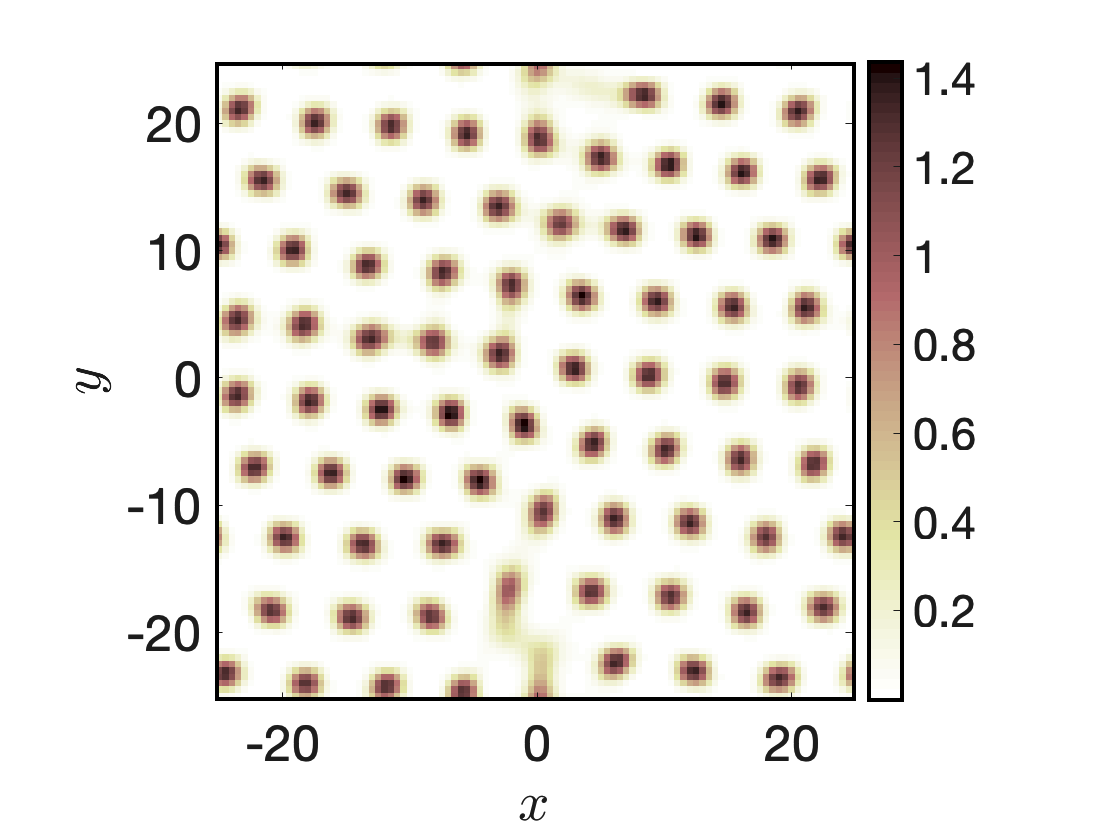}}
}
\hspace{-4pt}
\subfigure[pollutant $t=300$]{
    \includegraphics[width=0.31\textwidth]{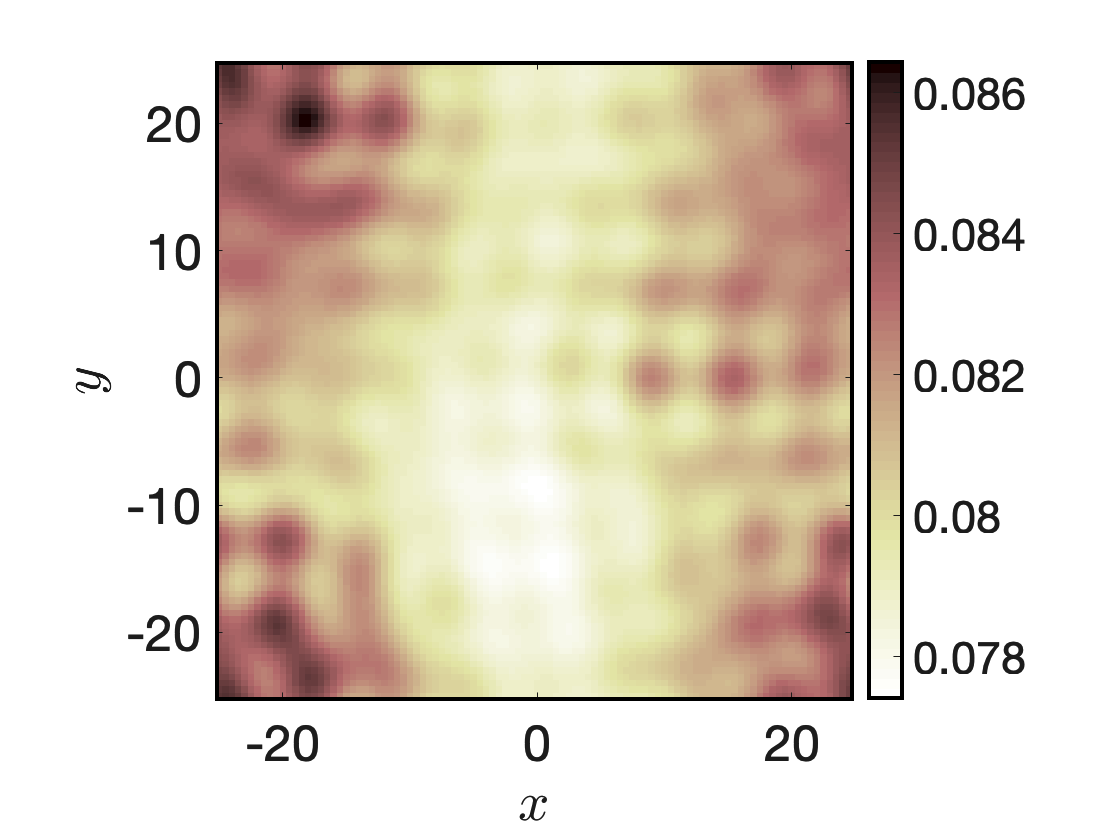}
}
\hspace{-4pt}
\subfigure[]{
    \includegraphics[width=6cm,height=5cm]{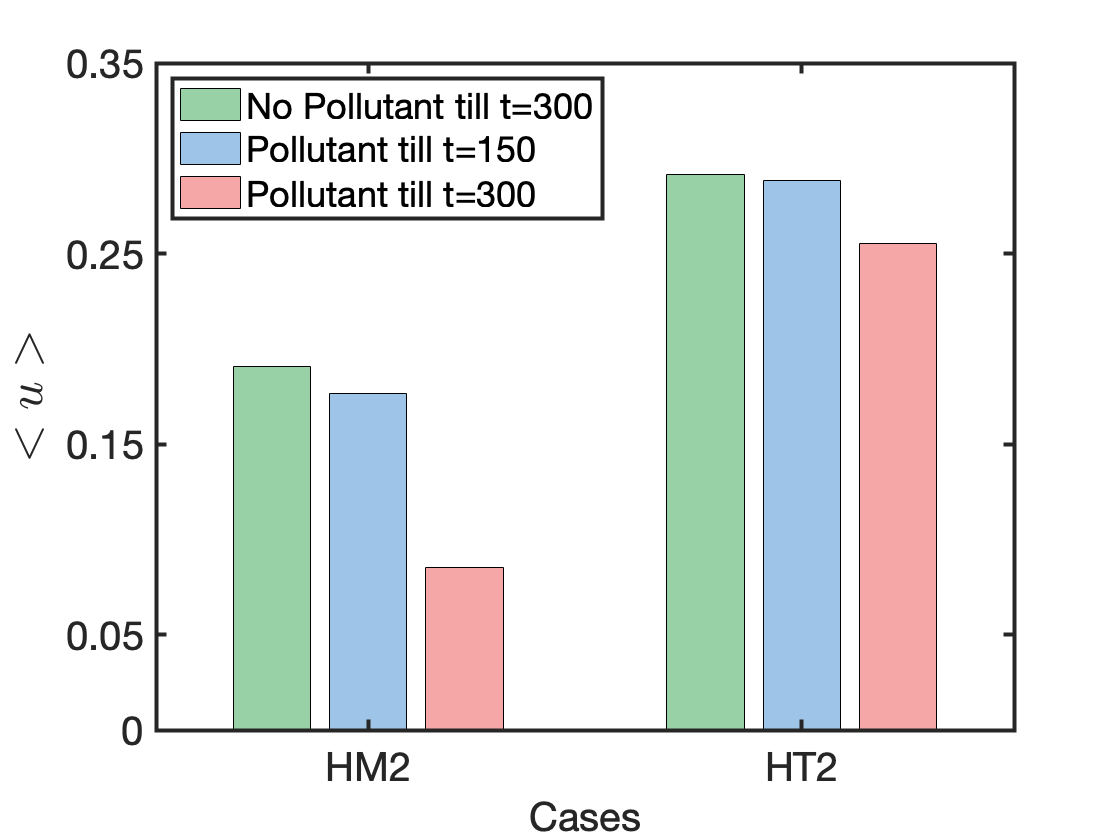}
} \hspace{-4pt}
\subfigure[]{
    \includegraphics[width=8.5cm,height=5cm]{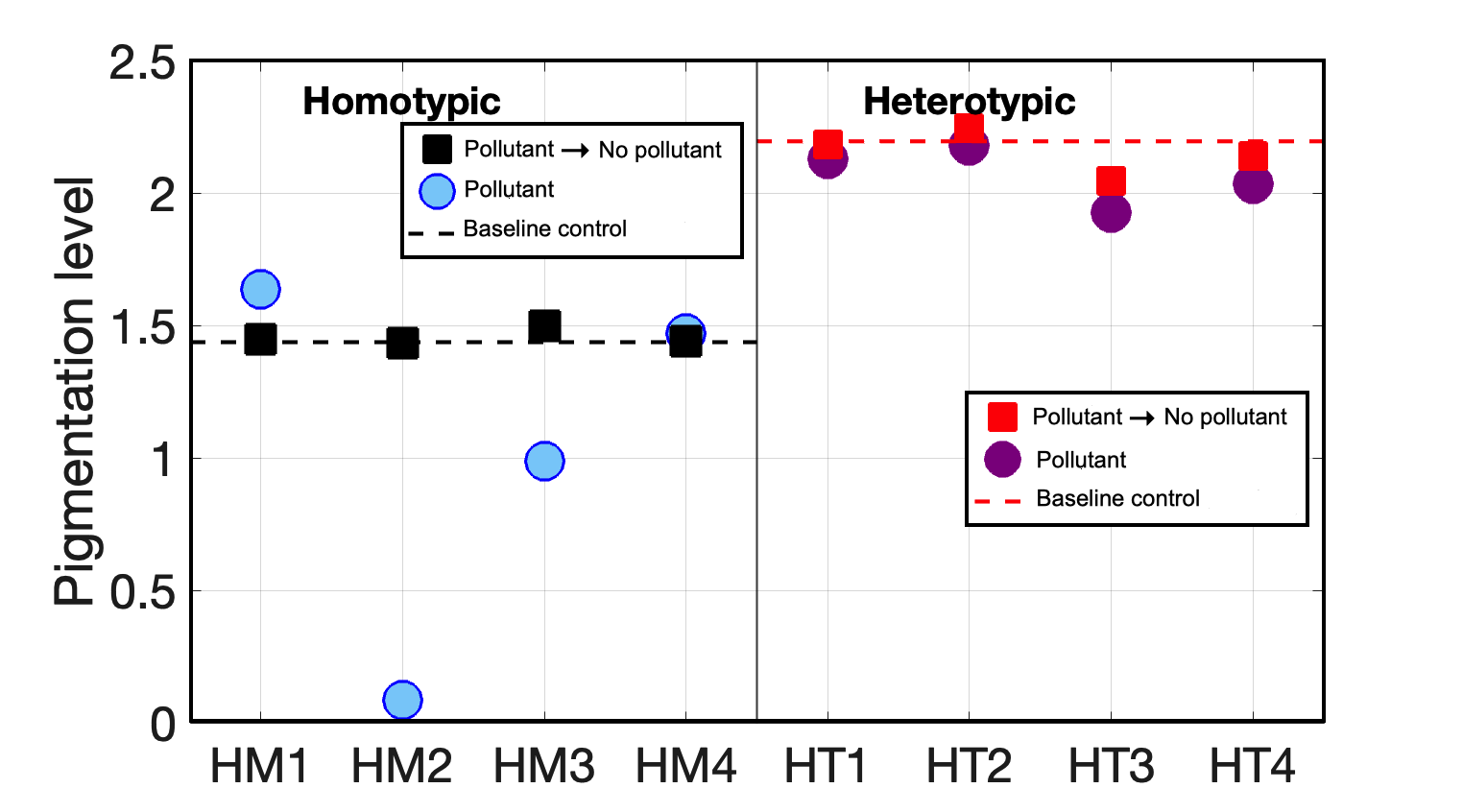}
    \label{fig:summary_delay}
}

\vspace{-6pt}
\caption{
Time-dependent effects of pollutant exposure in the homotypic case HM2.  
(a) Spatial pattern at $t=150$ under pollutant exposure.  
(b) Partial recovery of the pattern at $t=300$ after pollutant removal.  
(c) Strong suppression of pattern formation under continuous pollutant exposure at $t=300$.  
(d) Mean melanophore density ($<u>$) for HM2 and HT2 under three pollutant exposure regimes.  
(e) Summary of pigmentation changes for homotypic and heterotypic cases.
}
\label{fig:time_dependent_toxin}
\end{figure}

In this section, we examine how pattern formation and melanophore density change when fish experience pollutant exposure only during a short developmental stage, compared with continuous long-term exposure. We consider three treatment regimes: (i) a \emph{short-stage exposure} treatment, in which pollution is applied during the early phase (time \(0\text{--}150\)) and removed thereafter (time \(150\text{--}300\)); (ii) a \emph{continuous exposure} treatment, in which pollution is present throughout the entire simulation; and (iii) a \emph{control} treatment with no pollutant at any time. Among all scenarios examined, the HM2 case exhibited the greatest loss of pigmentation in the homotypic setting. In contrast, for the heterotypic case, the pigmentation level did not show a significant difference between the short-stage and continuous exposure treatments. A full comparison of the three exposure regimes is presented in Fig.~\ref{fig:time_dependent_toxin}.

In the first treatment, the zebrafish is exposed to the pollutant only during a short-term window (until $t = 150$) and then transferred back to fresh water. At $t = 150$, we observe a faded spot pattern (cf.~Fig.~\ref{fig:time_dependent_toxin}(a)), indicating that the pollutant has disrupted pattern maturation. The melanophore density is reduced, and the spatial pattern is less developed compared with the control. After removal from the polluted environment, the system partially recovers (cf.~Fig.~\ref{fig:time_dependent_toxin}(b)). The melanophore density increases and well-defined spots re-emerge; however, the maximum density remains slightly lower than in the control (cf.~Fig.~\ref{fig:time_dependent_toxin}(d)). This suggests that short-term exposure can hinder normal pattern development, although partial recovery is possible once the environment improves. In contrast, when the zebrafish is exposed continuously from $t = 0$ to $t = 300$, we observe the strongest suppression of formation of pattern (cf.~Fig.~\ref{fig:time_dependent_toxin}(c)). Under both interaction mechanisms, HM2 and HT2, the model~\eqref{mod:Movement_model} produces the most severe hypopigmentation by $t = 300$ (cf.~Fig.~\ref{fig:time_dependent_toxin}(d)). This indicates that continuous pollutant exposure exerts the greatest detrimental effect on pigment pattern formation. We summarize the results in Fig.~\ref{fig:time_dependent_toxin}(e)).





\begin{figure}[ht!]
\centering
\footnotesize
\renewcommand{\arraystretch}{1}

\includegraphics[width=0.45\textwidth, trim=5 0 5 5, clip]{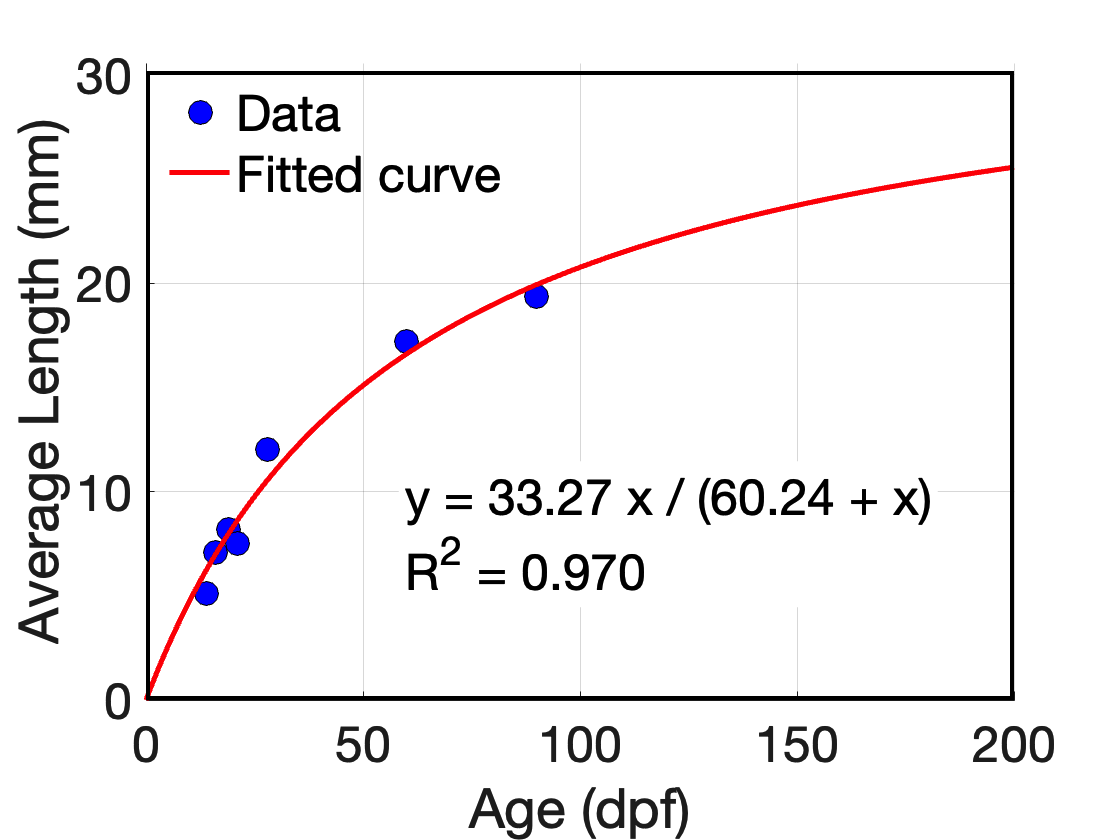}

\vspace{-5pt}
\caption{Fitted average domain length increase over age (days post fertilization, data from \cite{wallace2018effectiveness});  
}
\label{fig:length_fitting}
\end{figure}


\newcommand{\wA}{0.16\textwidth} 
\newcommand{\wB}{0.18\textwidth} 
\newcommand{\wC}{0.20\textwidth} 
\newcommand{\wD}{0.22\textwidth} 
\newcommand{\wE}{0.23\textwidth} 


\subsection{Hypo- and hyperpigmentation in pattern formation during fish development}

\begin{figure}[ht!]
\centering
\scriptsize
\setlength{\tabcolsep}{0pt}   
\renewcommand{\arraystretch}{0}


\subfigure[8 dpf]{\includegraphics[width=\wA, trim=5 5 5 5, clip]{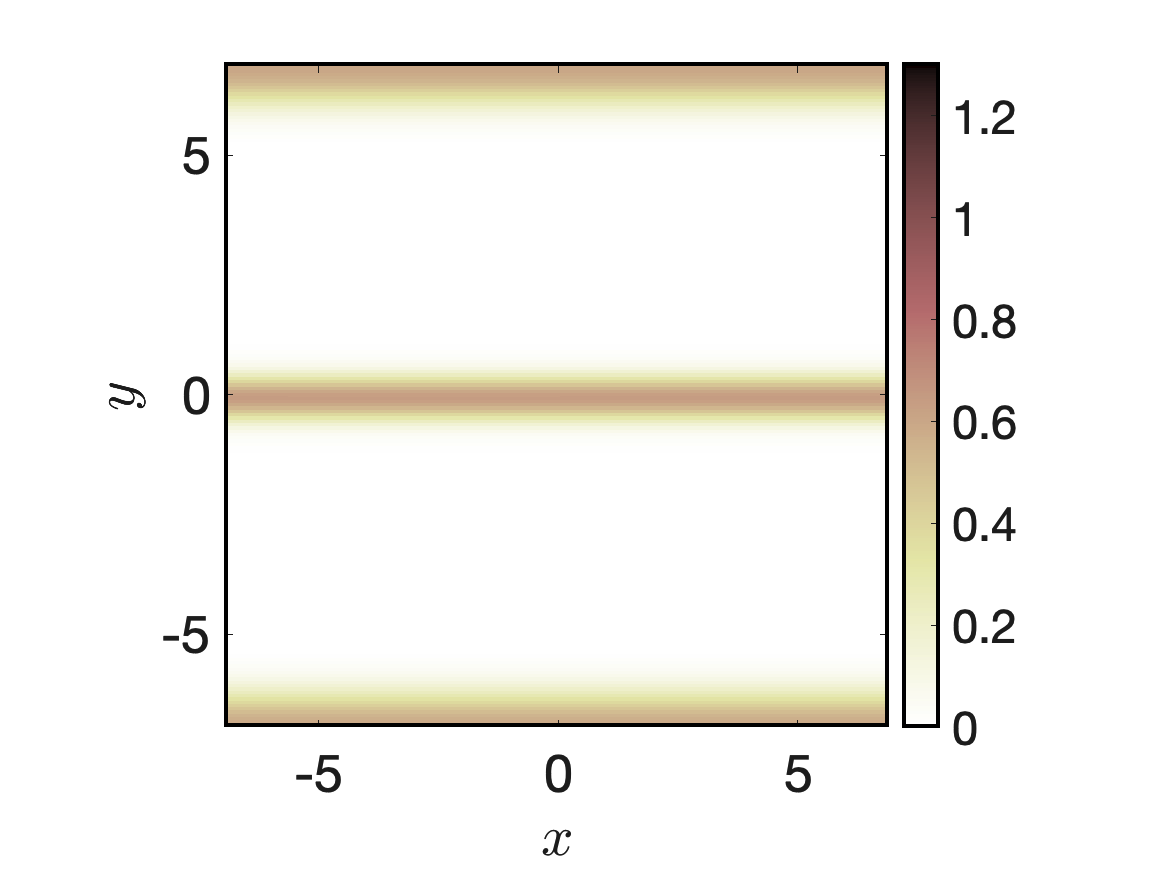}}\hspace{-10pt}
\subfigure[34 dpf]{\includegraphics[width=\wB, trim=5 5 5 5, clip]{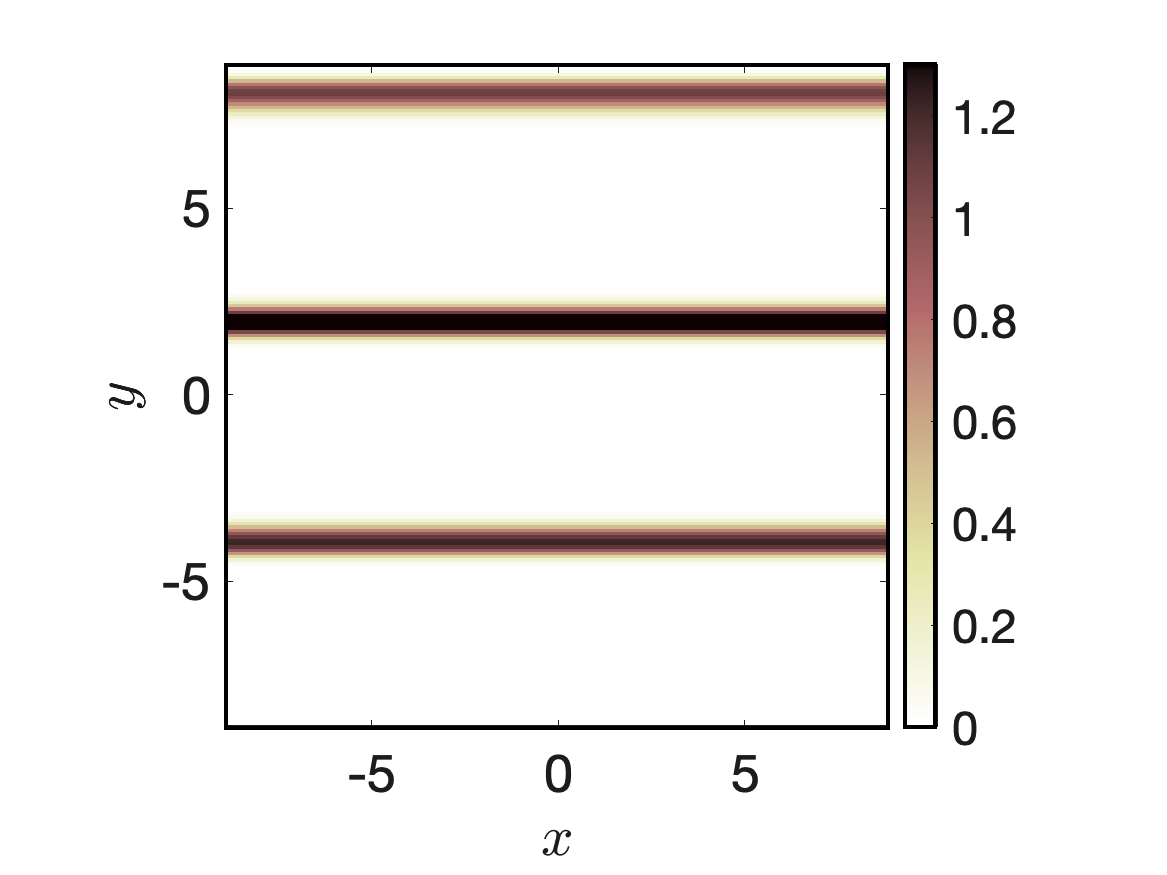}}\hspace{-10pt}
\subfigure[86 dpf]{\includegraphics[width=\wC, trim=5 5 5 5, clip]{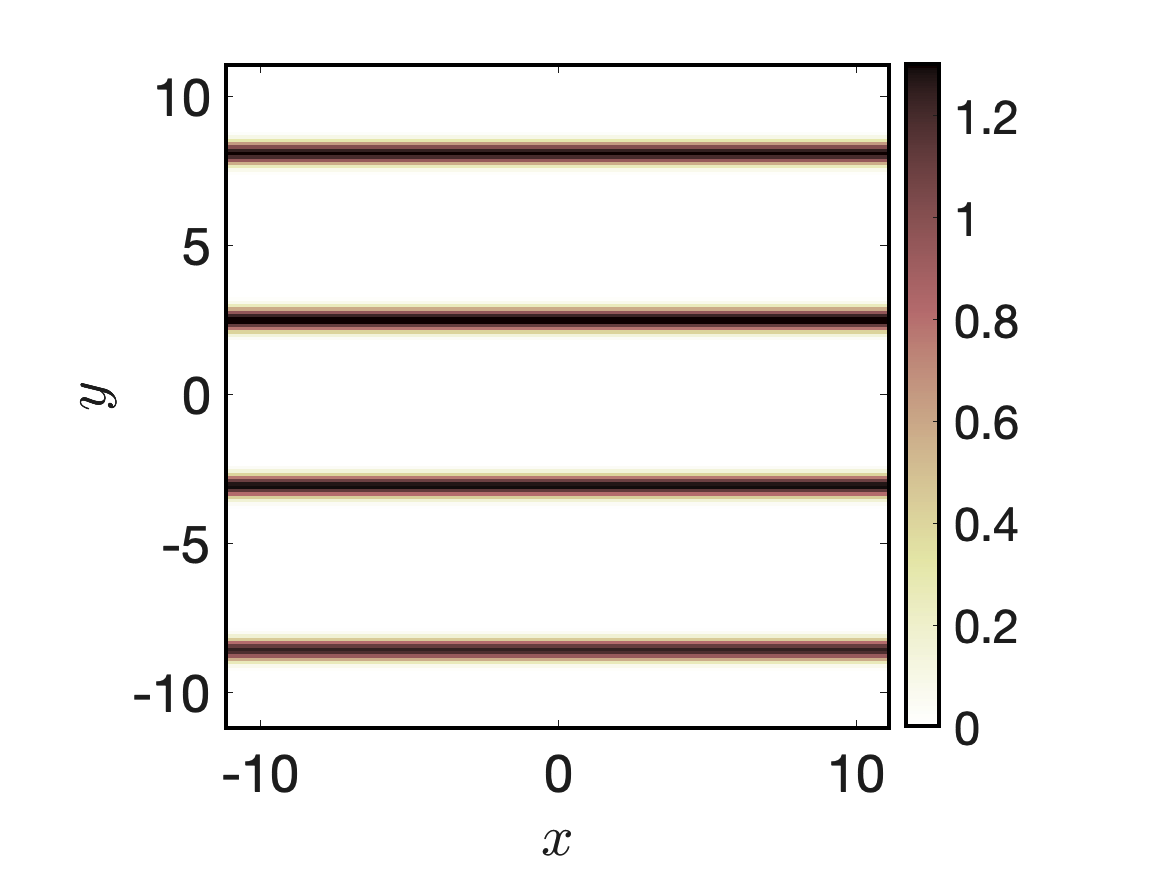}}\hspace{-10pt}
\subfigure[182 dpf]{\includegraphics[width=\wD, trim=5 5 5 5, clip]{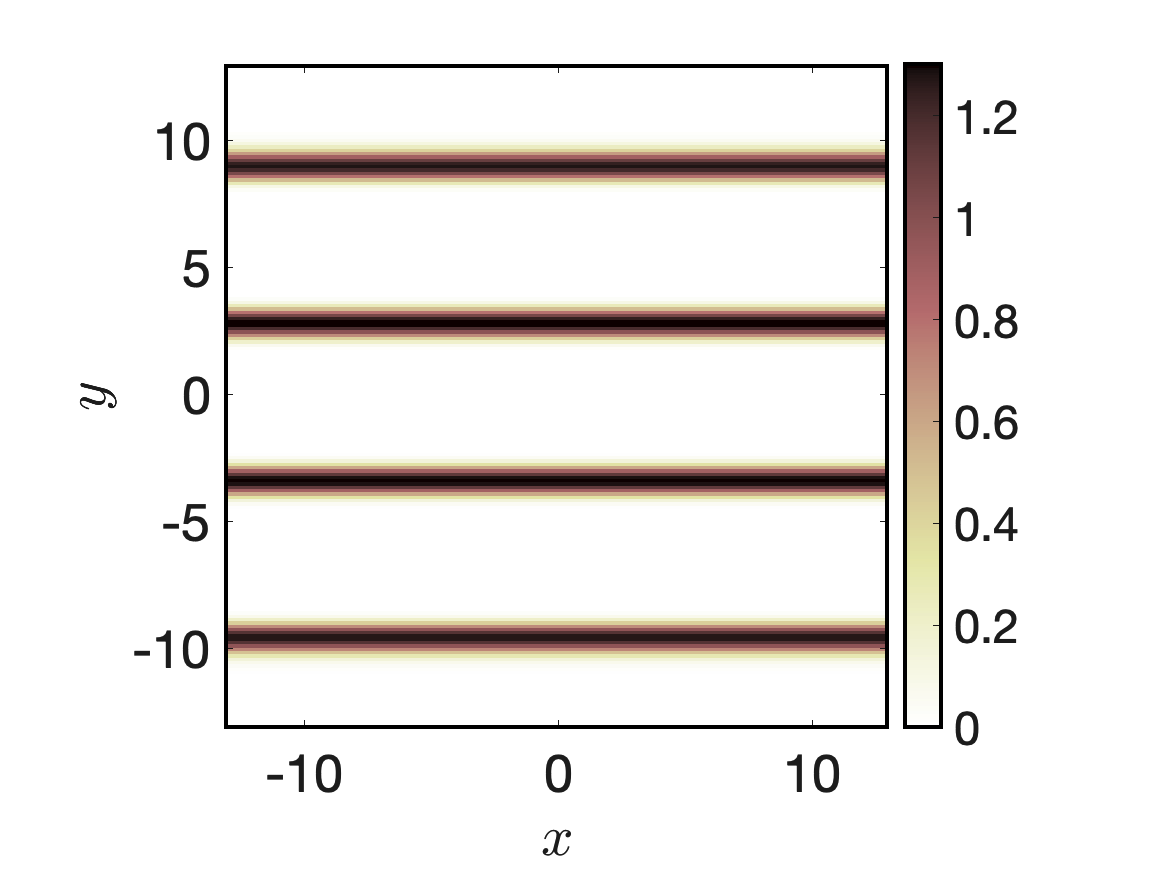}}\hspace{-10pt}
\subfigure[251 dpf]{\includegraphics[width=\wE, trim=5 5 5 5, clip]{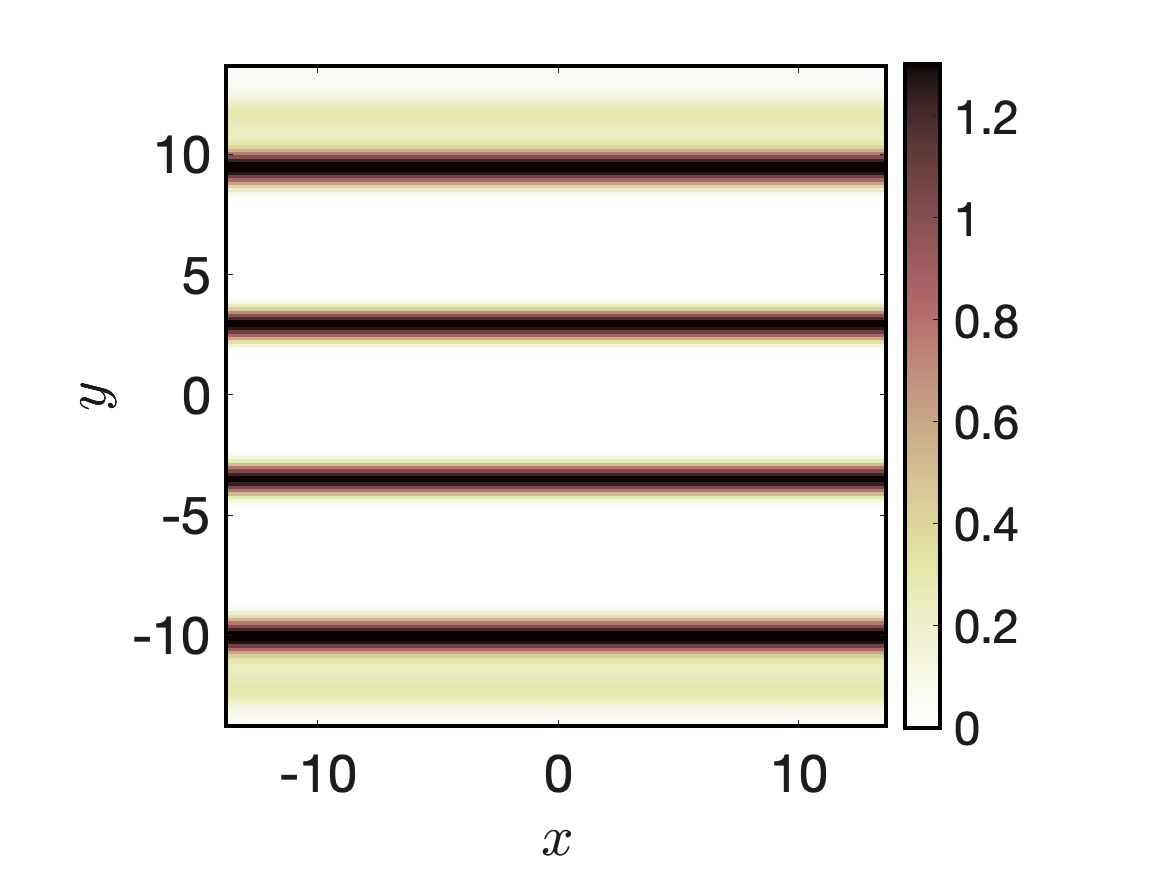}}\\[-7pt]
\caption{Simulation results showing pattern changes during growth. 
As the domain expands, the number of stripes increases. 
 Parameters: $R_{vv}=3$, $A_{uu}=5$, $A_{vv}=4$, $d_{uu}=5$, $d_{vv}=3$, $D_{uu}=D_{vv}=0.8$, and domain growth rate $\tfrac{33.27}{60.24+t}$. }
\label{fig: growning_domain_strip_without_toxin}
\end{figure}

To account for the effect of fish body growth on pattern formation, we model the fish length as a one-dimensional domain that expands over time. We consider a strip pattern as an example. The average fish body length growth across ages is shown in 
Figure~\ref{fig:length_fitting}, and can be well approximated by function
\[
y(t) = \frac{c\,t}{a + t},
\]
with fitted parameters \( c = 33.27 \) and \( a = 60.24 \). This function captures a typical saturating growth, where the length increases rapidly during early development and gradually levels off as the fish approaches maturity. Considering the initial body length be \(L_0\), then at time \(t\), the domain length of the fish body is given by
\[
L(t)=\frac{c\,(t+t_0)}{a+t+t_0},
\]
where \(t_0=\dfrac{L_0 a}{c-L_0}\) ensures \(L(0)=L_0\). This formulation allows the spatial domain to evolve smoothly in time. As the domain expands, pattern formation also evolves: stripe spacing increases, and new stripes may emerge as the fish continues to grow. This effect is illustrated in Fig.~\ref{fig: growning_domain_strip_without_toxin}, which models stripe formation on a growing fish body in a local region.

Mathematically, during development, the position of each cell can be tracked over time. 
We define the mapping
\begin{equation}
    \Gamma(X,t) = \frac{X}{L_0}\,
    \frac{c(t + t_0)}{a + t + t_0},
\end{equation}
which satisfies the initial and boundary conditions
\begin{equation}
    \Gamma(X,0) = X, \qquad \Gamma(0,t) = 0.
\end{equation}
Here, $\Gamma(X,t)$ represents the position at time $t$ of a cell that was initially located at $x = X$. 
This mapping explicitly describes the trajectory of individual cells under domain growth and allows growth-induced advection to be incorporated into the governing equations.

We introduce the rescaled coordinate
\[
\xi = \frac{x}{r(t)} \in [-1, 1],
\]
where
\begin{equation}
    \label{eq:r}
    r(t) = \Gamma(1,t) = 
    \frac{c(t + t_0)}{L_0(a + t + t_0)}.
\end{equation} This transformation allows us to map the dynamics of cells in the growing domain onto a fixed reference domain $[-1,1]$ homomorphically.
The function \(r(t)\) represents the normalized moving boundary of the growing domain. It expands according to a half-saturation dynamics. In the rescaled coordinates, the transformed system for the pattern-forming variables \(u(\xi,t)\) and \(v(\xi,t)\) becomes
\begin{equation}
\label{mod:Movement_model_growing_domain}
\footnotesize
\begin{aligned}
u_t &= 
\frac{D_u}{r(t)^2} \Delta_{\xi\xi} u 
- \frac{1}{r(t)^2} 
\nabla_{\xi} \!\cdot\! \bigl(u \nabla_{\xi} (G_{uv} * v)\bigr)
- \frac{1}{r(t)^2} 
\nabla_{\xi} \!\cdot\! \bigl(u \nabla_{\xi} (G_{uu} * u)\bigr)
+ \chi(\xi,t)\,\nabla_{\xi} u 
 + f(u,v), \\[4pt]
v_t &= 
\frac{D_v}{r(t)^2} \Delta_{\xi\xi} v 
- \frac{1}{r(t)^2} 
\nabla_{\xi} \!\cdot\! \bigl(v \nabla_{\xi} (G_{vu} * u)\bigr)
- \frac{1}{r(t)^2} 
\nabla_{\xi} \!\cdot\! \bigl(v \nabla_{\xi} (G_{vv} * v)\bigr)
+ \chi(\xi,t)\,\nabla_{\xi} v 
 + g(u,v),
\end{aligned}
\end{equation}
where
\[
\chi(\xi,t) = \frac{\xi\,\dot{r}(t)}{L_0\,r(t)},
\]
and \(\dot{r}(t)\) is the time derivative of \(r(t)\). The term \(\chi(\xi,t)\,\nabla_{\xi} u\) accounts for advection due to the stretching of the domain. Periodic boundary conditions are imposed on the rescaled domain \(\xi \in [-1,1]\):
\begin{equation}
    u(-1,t) = u(1,t), \quad 
    \partial_{\xi} u(-1,t) = \partial_{\xi} u(1,t), \quad
    v(-1,t) = v(1,t), \quad 
    \partial_{\xi} v(-1,t) = \partial_{\xi} v(1,t).
\end{equation}
These conditions indicate that cells can move continuously across the surface, allowing flux between any local domain within the fish body.

As the spatial domain grows at the rate $\frac{ct}{a+t}$, with $c=33.27$ and $a=60.24$, we simulate our model to explore pattern formation in the absence of external pollutant effects (cf. Fig.~\ref{fig: growning_domain_strip_without_toxin}). We examine the evolving patterns at several time snapshots $t = 8, 34, 86, 182,$ and $251$. Starting from a configuration with a single central stripe and two boundary stripes, additional stripes progressively emerge during development. As growth proceeds, pigmentation intensifies, and the fully developed adult pattern exhibits darker coloration with well-defined stripe boundaries.
\begin{figure}[ht!]
\centering
\scriptsize
\setlength{\tabcolsep}{0pt}
\renewcommand{\arraystretch}{0}

\subfigure[No pollutant]{
    \includegraphics[width=0.25\textwidth, trim=5 5 5 5, clip]{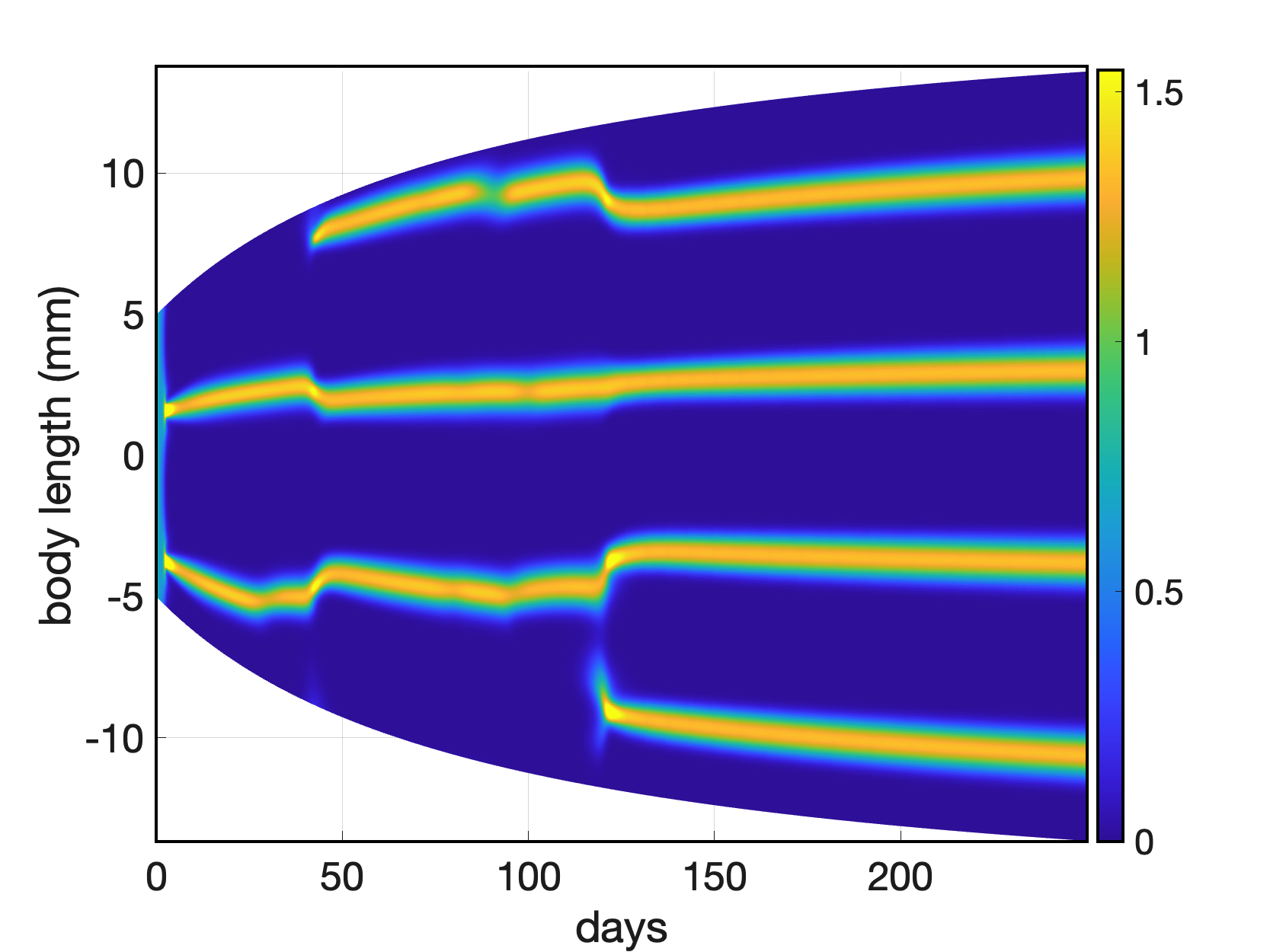}
}\hspace{-6pt}
\subfigure[HM1]{
    \includegraphics[width=0.25\textwidth, trim=5 5 5 5, clip]{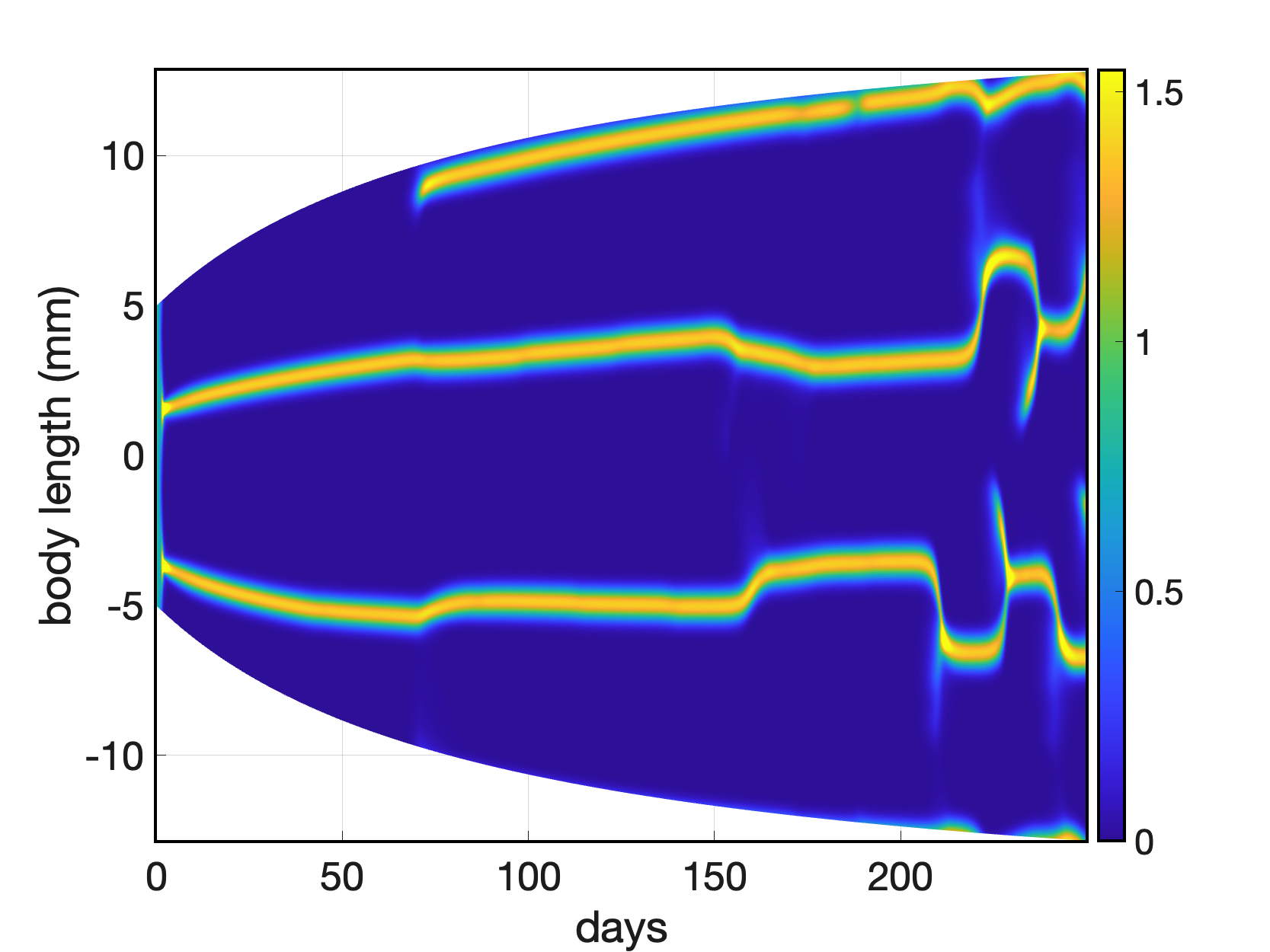}
}\hspace{-6pt}
\subfigure[HM2]{
    \includegraphics[width=0.25\textwidth, trim=5 5 5 5, clip]{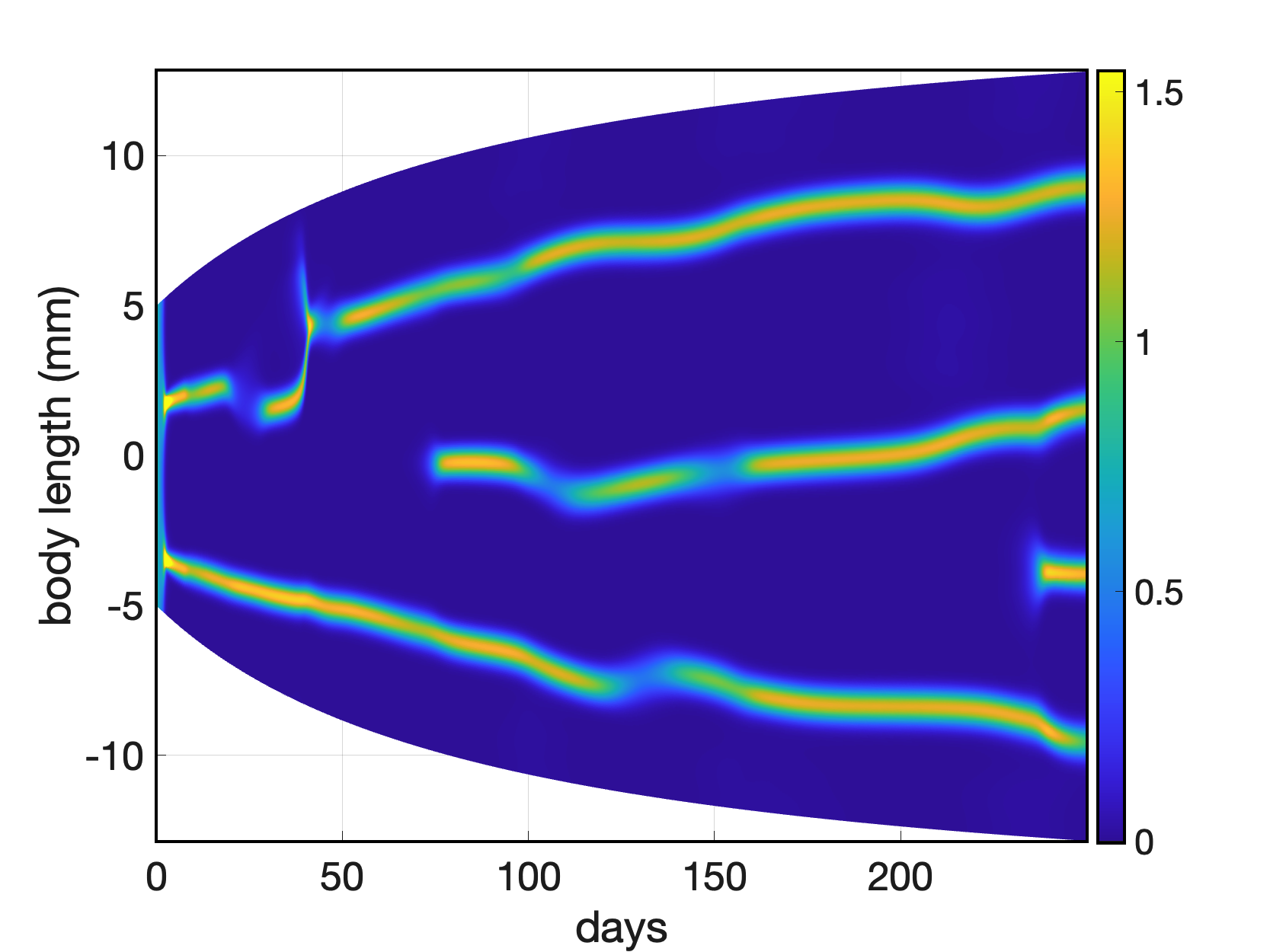}
}\hspace{-6pt}
\subfigure[HM3]{
    \includegraphics[width=0.25\textwidth, trim=5 5 5 5, clip]{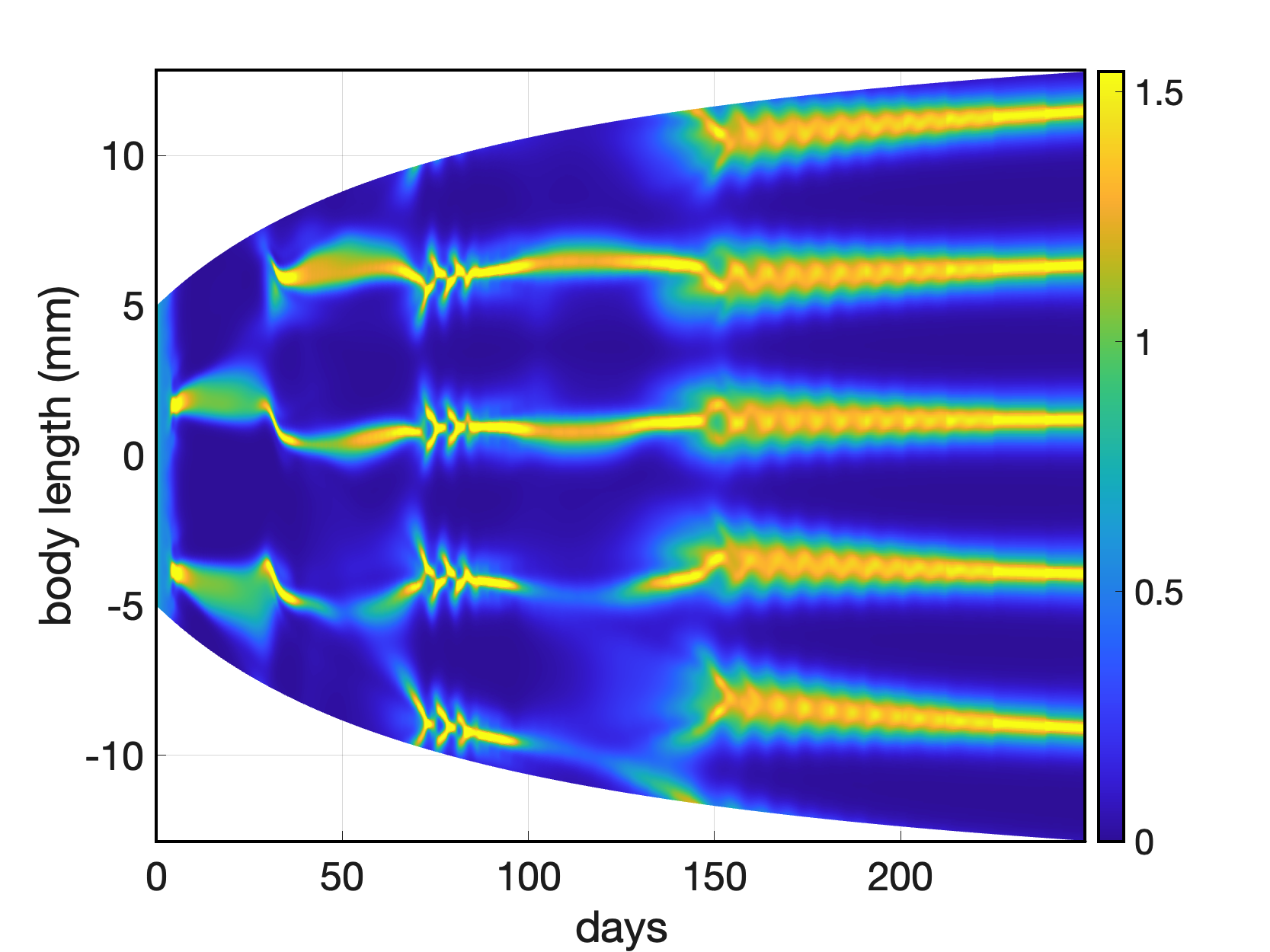}
}\hspace{-6pt}
\subfigure[HM4]{
    \includegraphics[width=0.25\textwidth, trim=5 5 5 5, clip]{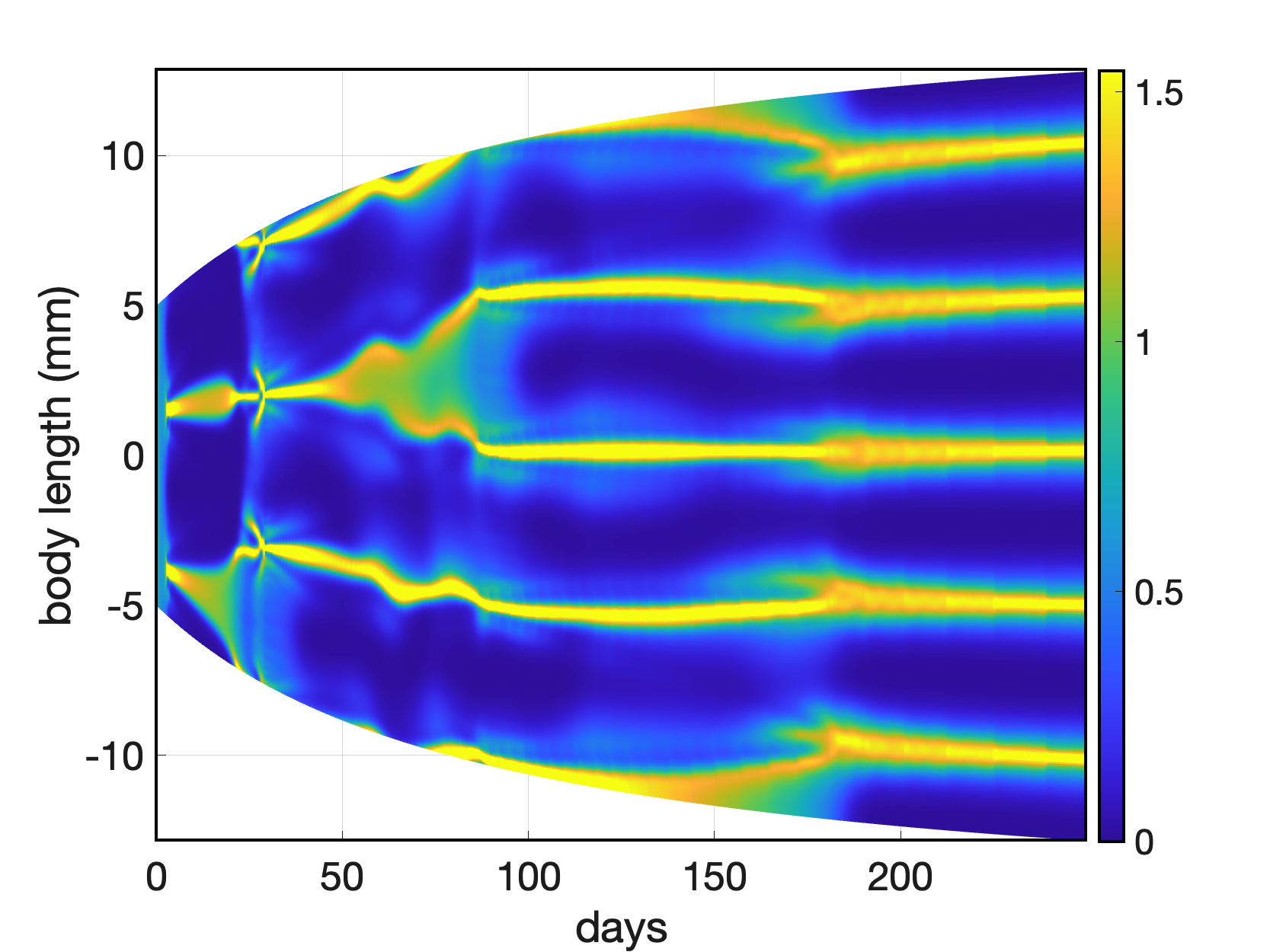}
}\\[-4pt]

\caption{
Comparison of observed and simulated pigment patterns during domain growth across the five hypotheses (No pollutant, HM1–HM4).
Shown here are representative time-dependent stripe-formation processes.
As the domain expands, additional stripes progressively emerge in both the experimental observations and the simulation results.
Parameters: $R_{vv}=3$, $A_{uu}=5$, $A_{vv}=4$, $d_{uu}=5$, $d_{vv}=3$, $D_{uu}=D_{vv}=0.8$, 
domain growth rate $33.27/(60.24+t)$.
}
\label{fig:growning_domain_stripes}
\end{figure}
We next investigate pollutant-induced changes during growth. To this end, we simulate the different cases described in Section~\ref{sec:Pollutant-induced changes in homotypic and heterotypic cell interactions} on a one-dimensional domain, focusing particularly on perturbations of homotypic interactions (HM1–HM4). Since homotypic interactions are much more sensitive to pollutants, we restrict our attention to these cases to better understand pollutant effects during development.  

Starting from a fully developed pattern at $t=0$ under control conditions, stripe addition continues over time in the growing domain (Fig.~\ref{fig:growning_domain_stripes}(a)). When pollutant-induced alterations in cell–cell interactions are incorporated, clear deviations from the control pattern emerge.  

Under HM1, stripe formation exhibits a pronounced developmental delay: the third stripe appears at approximately $t=75$, substantially later than in the control case, where it forms around $t=50$. By $t\approx300$, the fourth stripe only begins to emerge weakly, representing the slowest stripe addition among all cases. These results suggest that, under HM1, stripe development in the presence of pollutants proceeds significantly more slowly than in the control, with pronounced temporal variation, particularly during $t\approx200$–$300$.  

In the HM2 case, stripe structures persist throughout development but grow at a relatively slower rate. The third and fourth stripes appear at approximately $t=70$ and $t=270$, respectively, faster than in HM1 but still delayed relative to the control. Notably, the overall pigmentation intensity is markedly reduced, reflecting strong hypopigmentation.  

For HM3 and HM4, the pigment pattern undergoes substantial changes during early developmental stages ($t\lesssim100$), corresponding to rapid transitions during the larval-to-adult period. At later times, stripe patterns become more stable, with narrower stripes and a clear stabilization of spatial organization for $t\gtrsim200$. Across all four cases, the transient dynamics indicate that stripe growth rates differ markedly depending on which homotypic interactions are perturbed by pollutant exposure. In particular, HM3 and HM4 exhibit accelerated stripe growth, with the third, fourth, and fifth stripes appearing earlier than in the control case. This accelerated stripe differentiation process is clearly visible in Fig.~\ref{fig:growning_domain_stripes}(d,e). Among the four scenarios, only HM2 results in a sustained reduction of pigmentation intensity.

\section{Discussion}
 Developmental abnormalities induced by waterborne contaminants are widely used as biomarkers of aquatic ecosystem health \cite{mubashshir2025melanophore}. In this study, we present a quantitative and mechanistic framework to investigate how environmentally relevant water pollutants affect pigment pattern formation in fish. Pattern development relies on the coordinated behavior of pigment cells mediated by specific cellular interactions and morphogenetic processes \cite{patterson2019zebrafish}. For example, melanophores, xanthophores, and iridophores interact to generate the characteristic black-and-yellow stripe pattern observed in wild-type fish \cite{Volkening2015,Nakamasu2009,kondo2010reaction}.

Motivated by numerous pollutant-related empirical observations, we employed a nonlocal advection–diffusion model incorporating pollutant effects to investigate how altered pigment cell–cell interactions influence spatial patterning. Our results indicate that pollutant exposure not only alters overall pigmentation intensity, leading to hypo- or hyperpigmentation, but also reorganizes the spatial distribution of melanophores and xanthophores, resulting in increased pattern fragmentation and modified stripe configurations. These modeling predictions are consistent with our experimental findings: using ImageJ \cite{schneider2012nih}, we quantified pigment loss and disruptions in melanophore aggregation in methane-exposed fish, providing direct evidence of pollutant-induced hypopigmentation.

We extended the nonlocal reaction–diffusion–advection framework \cite{Volkening2015} to model interactions between melanophores and xanthophores in polluted environments. Polluted waters often contain complex mixtures of contaminants, including DIYA, fungicides, A$\alpha$C, and methane in oxygen-deficient conditions. Pollutant effects were modeled as modifications to adhesive and repulsive forces governing cell movement, incorporated through a differentiable interaction kernel that accounts for both short- and long-range interactions. Unlike classical diffusion, which assumes local random movement, the Morse-type kernel captures directed migration driven by attraction or repulsion over finite spatial ranges.

Numerical simulations revealed that pollutant-induced alterations in cell motility alone are sufficient to generate markedly different pigmentation patterns. Even without direct effects on reaction kinetics, changes in cell–cell interactions can fundamentally reshape pattern formation. Notably, homotypic interactions were far more sensitive to pollutant exposure than heterotypic interactions.

By systematically varying adhesion and repulsion parameters, we reproduced spot-like mutant patterns and quantified how specific interaction changes produce hypo- or hyperpigmentation. For example, decreasing short-range repulsion while increasing melanophore adhesion (HM1) led to cell aggregation and reduced tissue-wide melanin coverage, producing pronounced hypopigmentation. In contrast, higher repulsion among pigment cells preserved spot structures, and certain combinations of adhesive–repulsive interactions among xanthophores generated stripes with higher melanin density. These findings identify the particular homotypic interactions that drive either darker or lighter pigmentation outcomes. For heterotypic interactions, keeping homotypic interactions unchanged while modifying only the melanophore–xanthophore interactions produced mixed spot–stripe patterns. Pigment loss was most pronounced under conditions of strong intercellular repulsion. 

Our simulations also show that the timing and duration of pollutant exposure critically influence pigmentation outcomes. Short-term exposure delays pattern development, with partial recovery possible if conditions improve. In contrast, prolonged exposure results in severe hypopigmentation, for which recovery is minimal. Furthermore, we found that as pollutant levels increase, the overall effect generally becomes stronger; however, beyond a certain concentration, the effect saturates and does not intensify further. This phenomenon has been observed for specific pollutants, such as methane in our experiments, as well as A$\alpha$C and 2,4-DNT in previous studies \cite{Schmidt2016}. Our results suggest that it may represent a more general effect of environmental pollutants on pigment pattern formation.

To capture developmental dynamics, we incorporated a growing domain representing fish body elongation. Using empirically fitted zebrafish growth curves, we simulated stripe formation on an expanding domain. Stripe addition continues during growth, but pollutant-induced changes in homotypic interactions alter the rate of stripe formation: HM1 and HM2 slow stripe addition, whereas HM3 and HM4 accelerate it. This demonstrates that pollutant effects on cell–cell interactions can modulate both spatial patterning and temporal dynamics during development.

In summary, our work highlights the importance of understanding how pollutant-induced changes in short- and long-range cell–cell interactions shape pigment pattern formation. While direct experimental quantification of these interaction parameters remains challenging, our results provide a strong rationale for future experiments to investigate how contaminants modulate adhesive and repulsive forces between chromatophores. Such studies will be critical for precisely determining the underlying interaction mechanisms and for elucidating their role in driving pollutant-induced pigment pattern changes.

 \section*{Authors Contribution}

P.R.C: Conceptualization, Methodology, Data analysis, Formal Analysis, Investigation, Visualization, Writing - Original Draft, Writing - Review \& Editing; T.X.W: Conceptualization, Methodology, Formal Analysis, Investigation, Visualization, Writing - Original Draft, Writing - Review \& Editing.
A.M.C: Laboratory experiment, Data collection.
K.B.T: Conceptualization, Resources, Funding acquisition, Writing - Review \& Editing, Supervision.
H.W: Conceptualization, Methodology, Project Administration, Validation, Funding acquisition, Writing - Review \& Editing, Supervision.

\section*{Acknowledgments}
This research was partially supported by an NSERC Alliance Missions grant on greenhouse gas research. The research of Hao Wang was partially supported by the Natural Sciences and Engineering Research Council of Canada (Individual Discovery Grant RGPIN-2025-05734) and the Canada Research Chairs program (Tier 1 Canada
Research Chair Award).

\section*{Conflict of interest}
The authors declare that there are no conflicts of interest.

\bibliographystyle{plain}   
\bibliography{reference}

\end{document}